\documentclass[twocolumn,a4paper,aps,pre]{revtex4}
\usepackage{amsmath,amssymb}
\usepackage{graphics}

\begin{document}

\title{Mode-coupling theory for the slow collective dynamics of fluids
adsorbed in disordered porous media}

\author{V. Krakoviack}

\affiliation{Laboratoire de Chimie, UMR CNRS 5182, {\'E}cole Normale
Sup{\'e}rieure de Lyon, 46 All{\'e}e d'Italie, 69364 Lyon cedex 07,
France}

\date{\today}

\begin{abstract}
We derive a mode-coupling theory for the slow dynamics of fluids
confined in disordered porous media represented by spherical particles
randomly placed in space.  Its equations display the usual nonlinear
structure met in this theoretical framework, except for a linear
contribution to the memory kernel which adds to the usual quadratic
term. The coupling coefficients involve structural quantities which
are specific of fluids evolving in random environments and have
expressions which are consistent with those found in related problems.
Numerical solutions for two simple models with pure hard core
interactions lead to the prediction of a variety of glass transition
scenarios, which are either continuous or discontinuous and include
the possibility of higher-order singularities and glass-glass
transitions. The main features of the dynamics in the two most generic
cases are reviewed and illustrated with detailed
computations. Moreover, a reentry phenomenon is predicted in the low
fluid-high matrix density regime and is interpreted as the signature
of a decorrelation mechanism by fluid-fluid collisions competing with
the localization effect of the solid matrix.
\end{abstract}

\maketitle

\section{Introduction}

Over the last fifteen years, the dynamics of glassforming liquids
under nanoscale confinement has attracted much attention. A great
variety of molecular systems under many different types of confinement
has been subjected to virtually all available experimental techniques,
while at the same time extensive studies of simple model systems using
molecular dynamics simulations were undertaken
\cite{Confit2000,Confit2003,AlcMcK05JPCM,AlbCoaDosDudGubRadSli06JPCM}.

The main reason for this rapidly growing interest is that confinement
is considered as a potential tool to investigate the concept of
cooperativity, a key ingredient of many glass transition theories
\cite{Sil99JNCS,Edi00ARPC,Ric02JPCM}. Indeed, there are now many
evidences that the dynamics of deeply supercooled liquids is
inhomogeneous and that dynamically correlated groups of molecules play
a crucial role in the slowing-down of the dynamics when the
temperature is decreased. But, up to now, many questions pertaining,
for instance, to the shape, size or temperature evolution of these
dynamical heterogeneities remain essentially unanswered (for recent
progress, see, however,
Ref.~\cite{BerBirBouCipElMLHoLadPie05Science}).

In confinement, geometric constraints associated with the pore shape
are imposed to the adsorbed fluid and new characteristic length
scales, like the pore size, come into play. Thus, by looking for
alterations in the dynamics under confinement compared to the bulk,
one can hope to pinpoint some of the elusive characteristic features
of the dynamical heterogeneities. For instance, in the simplest
scenarios, deviations from the bulk behavior are expected to occur due
to finite size effects, when the typical size of the dynamical
heterogeneities in the bulk would become larger than the pore
size. Confinement effects would thus provide a ruler to measure
dynamical heterogeneities.

The situation actually turns out more complex. Indeed, a direct
comparison between the bulk and confined fluids is only meaningful if
the physical phenomena which are specific to confinement have a weak
impact on the properties of the imbibed fluid or at least if their
influence is sufficiently well understood that it can be corrected
for.  This is usually not so and strong confinement effects are often
observed, for instance, the formation of structured layers of almost
immobile molecules at the fluid-solid interface. So, in fact, dynamics
in confinement should be addressed as a problem of its own, not
necessarily with reference to the bulk.

The variety of the systems to consider is immense. Porous media can
differ in the size, shape and topology of their pore space. They can
be made of various materials or receive different surface treatments,
leading to a wide range of fluid-solid interactions which adds to the
already great variability of the intermolecular interactions met with
usual glassformers. Thus, owing to the complexity of the field, a
reasonable microscopic theory, able to catch at least some of these
aspects, could be very helpful. Indeed, applied to various models, it
would allow to explore thoroughly the phenomenology of confined
glassforming liquids and maybe to disentangle the contributions of the
different physical phenomena which interplay in these systems.

\begin{figure}
\includegraphics*{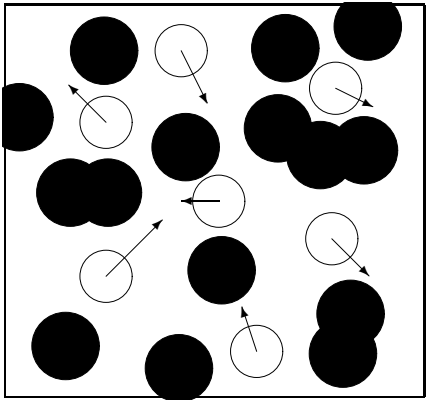}
\caption{\label{figsketch} Sketch of a QA system. In black, the
  immobile matrix particles. In white, with arrows symbolizing their
  movement, the fluid particles.}
\end{figure}

Many porous media, like controlled porous glasses and aerogels, are
disordered. In the past few years, a very useful and quite successful
model to deal with this kind of systems has been the so-called
``quenched-annealed'' (QA) binary mixture, first introduced by Madden
and Glandt \cite{MadGla88JSP,Mad92JCP}. In this model, sketched in
Fig.~\ref{figsketch}, the fluid molecules (the annealed component)
equilibrate in a matrix of particles frozen in a disordered
configuration sampled from a given probability distribution (the
quenched component).  The matrix is assumed to be statistically
homogeneous, so that, while for any single realization, the system
lacks translational and rotational invariance, all expectation values
computed with the matrix probability distribution will have the same
properties as in a truly translationally and rotationally invariant
system.  A common, but not unique, prescription is to take the
equilibrium distribution of some simple fluid system, so that the
various samples of the matrix can be thought of as the results of
instantaneous thermal quenches of this original equilibrium system,
hence the denomination ``quenched'' for the matrix component.

Thanks to the assumption of statistical homogeneity, as far as the
computation of matrix averaged quantities is concerned, QA mixtures
can be studied with great ease using simple extensions of standard
liquid state theoretical methods. This has been put to good use to
derive equations describing the structure and the thermodynamics of
these systems, either via diagrammatic techniques
\cite{MadGla88JSP,Mad92JCP,ForGla94JCP} or by application of the
replica trick
\cite{GivSte92JCP,LomGivSteWeiLev93PRE,GivSte94PA,RosTarSte94JCP}.

The aim of the present work is now to develop a dynamical theory for
QA mixtures, able to deal with the problem of the glass transition in
confinement. This will take the form of an extension to QA systems of
the ideal mode-coupling theory (MCT) for the liquid-glass transition
\cite{BenGotSjo84JPC,Leu84PRA,LesHouches,GotSjo92RPP,Got99JPCM}.  The
MCT occupies a central place in the study of the dynamics of
supercooled liquids in the bulk, as one of the very few available
microscopic theories in this field. It has well known deficiencies, in
particular the fact that its predicted sharp ergodicity breaking
transition, the so-called ideal glass transition, is always located in
the regime of weak to mild supercooling, rather far from the
calorimetric glass transition point. But, quite remarkably, it has
been able to correctly predict novel nontrivial relaxation patterns
which develop in systems like colloidal suspensions with short-ranged
attractions \cite{DawFofFucGotSciSpeTarVoiZac01PRE} or fluids of
symmetric dumbbells \cite{ChoGot02PRE}, when the parameters of these
models are varied. It seems thus sensible to turn to this theoretical
framework for a systematic investigation of confined glassforming
liquids, aiming at understanding, at least qualitatively, how the
microscopic details of the fluid-solid system impact its
dynamics. Moreover, it is encouraging that recent computer simulation
data for confined fluids could be interpreted with the universal
predictions of the MCT
\cite{GalRovSpo00PRL,GalPelRov02EL,SchKobBin04JPCB}, showing that the
mode-coupling glass transition scenario might indeed be of some
relevance in confinement as well. Note that Ref.~\cite{GalPelRov02EL}
precisely deals with a QA system, as do
Refs.~\cite{Kim03EL,ChaJagYet04PRE,MitErrTru06PRE}, where other
simulation studies are reported.

Theories related to the one to be derived have already been discussed
in the literature. The mode-coupling approach to the
diffusion-localization transition in the classical random Lorentz gas
\cite{GotLeuYip81PRA,Leu83PRA,Sza04EL} is of particular relevance,
since this system is actually a QA mixture taken in the limit of a
vanishing density of the annealed component. Besides, the present
theory will borrow some ideas from this approach. Other works have
dealt with the conductor-insulator transition of quantum fluids in
random potentials, originally neglecting the fluid-fluid interactions
\cite{Got78SSC,Got81PMB} which were later reintroduced by Thakur and
Neilson \cite{ThaNei96PRB}. They appear as special cases of the
present theory, with additional implicit and uncontrolled
approximations in the treatment of the effect of randomness on the
static fluid correlations \cite{QuantumNote}. Finally, a similar
nonlinear feedback mechanism has been derived for the freezing of a
polymer chain in a quenched random medium using the self-consistent
Hartree approximation for the Langevin dynamics of the system
\cite{MigRosVil03EPJB}.

The paper is organized as follows. In Sec.~II, the MCT equations for
the collective dynamics of a QA mixture are derived and discussed,
while Sec.~III explores the relation between the dynamics in a
self-induced glassy phase and in a quenched random environment. In
Sec.~IV, dynamical phase diagrams for two simple models are computed,
which show two basic types of ideal liquid-glass transitions discussed
in more details in Sec.~V. Section VI is devoted to concluding
remarks. Preliminary reports on the present work can be found in
Refs.~\cite{Kra05PRL} and \cite{Kra05JPCM}.

\section{Mode-coupling theory for quenched-annealed systems}

In this section, the mode-coupling equations for the QA binary mixture
are derived and discussed. But, before proceeding with the dynamical
theory, a few static quantities have to be defined.

As mentioned in the introduction, in a QA system, the disordered
porous medium is represented by a collection of $N_m$ rigorously
immobile point particles, randomly placed in a volume $V$ at positions
denoted by $\mathbf{s}_1, \mathbf{s}_2, \ldots, \mathbf{s}_{N_m}$,
according to a given probability distribution
$\mathcal{P}(\mathbf{s}_1, \mathbf{s}_2, \ldots, \mathbf{s}_{N_m})$
\cite{MadGla88JSP,Mad92JCP}. Its overall density is $n_m=N_m/V$ and
the Fourier components of its frozen microscopic density, or, in
short, its frozen density fluctuations, are given by
\begin{equation}
\rho^m_\mathbf{q}=\sum_{j=1}^{N_m} e^{i \mathbf{q} \mathbf{s}_j},
\end{equation}
where $\mathbf{q}$ denotes the wavevector. Their disorder-averaged
correlation functions, which, because of the assumed statistical
homogeneity, are diagonal in $\mathbf{q}$ and only depend on its
modulus $q$, define the matrix structure factor
\begin{equation}
S^{mm}_q = \frac{1}{N_m} \overline{\rho^m_\mathbf{q}
\rho^{m}_\mathbf{-q}},
\end{equation}
where $\overline{\cdots}$ denotes an average over the matrix
realizations.

The fluid component consists of $N_f$ point particles (density
$n_f=N_f/V$) of mass $m$, which equilibrate at a temperature $T$ in
the random potential energy landscape created by the frozen matrix
particles. As in the bulk, for the present theory, we will be
interested in the dynamics of the fluid density fluctuations
\begin{equation}
\rho^f_\mathbf{q}(t)=\sum_{j=1}^{N_f} e^{i \mathbf{q} \mathbf{r}_j(t)},
\end{equation}
where $\mathbf{r}_j(t)$ is the position of the fluid particle $j$ at
time $t$. At equal times, they allow to define the fluid structure
factor
\begin{equation}
S^{ff}_q= \frac{1}{N_f} \overline{\langle \rho^f_\mathbf{q}
\rho^{f}_\mathbf{-q} \rangle},
\end{equation}
and the fluid-matrix structure factor
\begin{equation}
S^{fm}_q = \frac{1}{\sqrt{N_f N_m}} \overline{\langle
\rho^f_\mathbf{q} \rho^{m}_\mathbf{-q} \rangle},
\end{equation}
where $\langle \cdots \rangle$ denotes a thermal average \emph{taken
for a given realization of the matrix}, the disorder average
$\overline{\cdots}$ being performed \emph{subsequently}. 

For an ordinary binary mixture, the knowledge of the above three
structure factors would be enough to fully characterize the structure
at the pair level. This is not the case for a QA mixture. Indeed,
because the matrix component is quenched, for any single realization,
the system lacks translational and rotational invariance. It results
that, at variance with a bulk fluid, non-zero average density
fluctuations exist at equilibrium, i.e., $\langle \rho^f_\mathbf{q}
\rangle \neq 0$. It is only after averaging over disorder that the
symmetry is restored, so that $\overline{\langle \rho^f_\mathbf{q}
\rangle} = 0$. Thus, one is led to consider relaxing and non-relaxing
fluid density fluctuations, corresponding to
$\delta\rho^f_\mathbf{q}(t) = \rho^f_\mathbf{q}(t) - \langle
\rho^f_\mathbf{q} \rangle$ and $\langle \rho^f_\mathbf{q} \rangle$,
respectively, and to define the connected fluid structure factor
\begin{equation} \label{Sconn}
S^{c}_q = \frac{1}{N_f} \overline{\langle \delta \rho^f_\mathbf{q}
\delta \rho^{f}_\mathbf{-q} \rangle},
\end{equation}
and the disconnected or blocked fluid structure factor
\begin{equation} \label{Sbloc}
S^{b}_q = \frac{1}{N_f} \overline{\langle \rho^f_\mathbf{q} \rangle
\langle \rho^{f}_\mathbf{-q} \rangle},
\end{equation}
such that $S^{ff}_q = S^{c}_q + S^{b}_q$. This splitting of the fluid
pair correlations is well known from the replica theory of QA systems,
where it leads to the peculiar structure of the so-called replica
Ornstein-Zernike (OZ) equations
\cite{GivSte92JCP,LomGivSteWeiLev93PRE,GivSte94PA,RosTarSte94JCP},
which are given in Appendix \ref{app.oz} for reference.

Non-zero average fluid density fluctuations at equilibrium mean that,
even without any dynamical ergodicity breaking, they will have
time-persistent correlations. Indeed, defining the normalized total
density fluctuation autocorrelation function
\begin{equation}
\phi^T_q(t)= \frac{\overline{ \langle \rho^f_\mathbf{q}(t)
\rho^f_\mathbf{-q}(0) \rangle}}{N_f S^{ff}_q},
\end{equation}
one expects using standard arguments that
\begin{equation}
\lim_{t\to\infty} \phi^T_q(t) = \frac{\overline{
\langle\rho^f_\mathbf{q}\rangle \langle\rho^{f}_\mathbf{-q} \rangle
}}{N_f S^{ff}_q} = \frac{S^{b}_q}{S^{ff}_q} > 0.
\end{equation}
This is a general consequence of the fact that the fluid evolves in an
inhomogeneous environment and we stress that this is a true static
phenomenon.

Usually, mode-coupling theories are derived assuming that the statics
of the problem is solved. For this reason, since the calculation of
the contribution from the blocked correlations is actually a static
problem, the theory has been developed using the relaxing part of the
fluid density fluctuations as the central dynamical variable. Attempts
to derive a dynamical theory starting from the full density
fluctuations have resulted in complicated equations, which appeared
unfaithful to the statics of the problem, and thus were abandoned.

The theory is derived using standard projection operator methods, as
shown in Ref.~\cite{LesHouches} for bulk systems.  As usual, in a
first step, one obtains a generalized Langevin equation for the time
evolution of the normalized connected autocorrelation function of the
density fluctuations
\begin{equation}
\phi_q(t)= \frac{\overline{ \langle
\delta \rho^f_\mathbf{q}(t) \delta \rho^f_\mathbf{-q}(0) \rangle}}{N_f
S^{c}_q}. 
\end{equation}
It is formally the same as for the bulk, i.e.,
\begin{equation}\label{eqlangevin}
\ddot{\phi}_{q}(t) + \Omega_{q}^2 \phi_{q}(t) + \Omega_{q}^2 \int_0^t
d\tau m_q(t-\tau) \dot{\phi}_{q}(\tau)=0,
\end{equation}
with initial conditions $\phi_q(0)=1$, $\dot{\phi}_q(0)=0$, and
\begin{equation} \label{freqmic}
\Omega_{q}^2=\frac{q^2 k_B T}{m S^{c}_q}.
\end{equation}

The second step involves the calculation of the slow decaying portion
of the memory kernel $m_q(t)$ with a mode-coupling approach. We will
assume that the slow dynamics is dominated by three types of quadratic
variables, which can be separated in two classes. In the first class,
we have variables quadratic in the relaxing density fluctuations,
$\delta \rho^f_\mathbf{k} \delta \rho^f_\mathbf{q-k}$, in close
analogy with bulk MCT \cite{BenGotSjo84JPC}. In the second class,
inspired by previous studies on the Lorentz gas
\cite{GotLeuYip81PRA,Leu83PRA,Sza04EL}, we consider variables
expressing couplings of the relaxing density fluctuations to the two
frozen density fluctuations present in the problem, $\delta
\rho^f_\mathbf{k} \rho^m_\mathbf{q-k}$ and $\delta \rho^f_\mathbf{k}
\langle \rho^f_\mathbf{q-k} \rangle$. The last variable was omitted in
all previous works, including a recent account of the present theory
\cite{Kra05PRL}, with consequences to be discussed below.

The calculation is outlined in Appendix \ref{app.derivation} and we
only quote the result here. It reads $m_q(t)=\Gamma_q \delta(t) +
m^{(\text{MC})}_q(t)$, where $\Gamma_q$ is a friction coefficient
associated with fast dynamical processes, and
\begin{equation} \label{kernel}
m^{(\text{MC})}_q(t) = \int \frac{d^3\mathbf{k}}{(2\pi)^3} \left[
V^{(2)}_{\mathbf{q},\mathbf{k}} \phi_{k}(t) \phi_{|\mathbf{q-k}|}(t) +
V^{(1)}_{\mathbf{q},\mathbf{k}} \phi_{k}(t) \right],
\end{equation}
with
\begin{subequations} \label{vertices}
\begin{equation}\label{vtwo}
V^{(2)}_{\mathbf{q},\mathbf{k}} = \frac{1}{2} n_f S^{c}_q
\left[\frac{\mathbf{q}\cdot\mathbf{k}}{q^2} \hat{c}^{c}_k +
\frac{\mathbf{q}\cdot(\mathbf{q-k})}{q^2}
\hat{c}^{c}_{|\mathbf{q-k}|}\right]^2 S^{c}_k
S^{c}_{|\mathbf{q-k}|},
\end{equation}
and
\begin{equation}\label{vone}
V^{(1)}_{\mathbf{q},\mathbf{k}} = n_f S^{c}_q \left[
\frac{\mathbf{q}\cdot\mathbf{k}}{q^2} \hat{c}^{c}_k +
\frac{\mathbf{q}\cdot(\mathbf{q-k})}{q^2} \frac{1}{n_f} \right]^2
S^{c}_k S^{b}_{|\mathbf{q-k}|},
\end{equation}
\end{subequations}
where $\hat{c}^{c}_q$, the Fourier transform of the connected direct
correlation function, has been introduced.

Based on these equations, a few general comments are in order. First
of all, the derived expressions retain the mathematical structure of
the typical mode-coupling equations which have been extensively
studied in Ref.~\cite{LesHouches}. Thus, all known properties of the
solutions of MCT equations, in particular near the transition, apply
in the case of the QA mixture. A significant addition compared to the
bulk is however the presence of a linear term in the memory function,
which opens the possibility of continuous ideal glass transitions.

Like in the theory for the bulk, which is recovered in the limit
$n_m\to0$, the slow dynamics is fully determined by smoothly varying
static quantities. Interestingly, no explicit reference to the matrix
is visible in the equations. Indeed, the only required information is
$n_f$, $S^{c}_q$, $S^{b}_q$, and $\hat{c}^{c}_q$, i.e., quantities
characterizing the fluid component of the QA system. These functions
and the relations between them are generically meaningful for the
description of fluids evolving in statistically homogeneous random
environments and they are by no means restricted to the model of the
QA mixture (for the case of non-particle-based random fields, see
Refs.~\cite{MenDas94PRL,KieRosTar99JSP}). Therefore, one can expect
that the present dynamical theory shares the same degree of generality
and is applicable in its present form out of the strict context of the
QA binary mixture. In fact, this shows up nicely in the process of
deriving the equations, since one finds that the contributions
resulting from the coupling of the random forces to $\delta
\rho^f_\mathbf{k} \rho^m_\mathbf{q-k}$ are identically zero (see
Appendix \ref{app.derivation}). Thus, it is enough to consider only
two types of quadratic variables, $\delta \rho^f_\mathbf{k} \delta
\rho^f_\mathbf{q-k}$ and $\delta \rho^f_\mathbf{k} \langle
\rho^f_\mathbf{q-k} \rangle$, precisely those which are not specific
to QA systems, to obtain the same dynamical equations. Further
evidence of the generic character of the present equations is given in
Appendix \ref{app.pspin}, where the dynamics of a mean-field
spin-glass model in a random magnetic field is shown to obey equations
with exactly the same structure.

Along the same line, it is noteworthy that the total fluid structure
factor $S^{ff}_q$ does not appear in the derived dynamical equations.
Only $S^{c}_q$ and $S^{b}_q$ do. It results that, in the framework of
the MCT, the global fluid correlations are of limited relevance to
discuss the relation between the statics and the dynamics. A much more
important aspect is rather the balance between the connected and
disconnected contributions to these correlations.  This result is very
welcome, since similar differences in the roles played by these three
types of correlations are known to be crucial in the physics of QA
systems. This is best illustrated by the compressibility sum rule
\cite{ForGla94JCP,RosTarSte94JCP}, which precisely involves $S^{c}_q$
and not $S^{ff}_q$, as one would naively expect from the relation for
the bulk. It is thus reassuring that this feature has been preserved
despite the uncontrolled approximations involved in the derivation of
the theory. Moreover, this finding allows one to clarify the issues
raised in Refs.~\cite{ChaJagYet04PRE,SchKobBin04JPCB} about the
possibility of a description of the dynamics in confinement with
theories based on structural quantities only. Indeed, it was observed
there that systems with identical global fluid correlations could have
significantly different dynamics, a result which appeared as a
challenge to such approaches. The present MCT shows that this is not
necessarily so and that one should also consider the connected and
blocked correlations before a conclusion can be reached.

About the form of the vertices, one can note the familiar expression
of $V^{(2)}_{\mathbf{q},\mathbf{k}}$, which measures the coupling of
the random forces to $\delta \rho^f_\mathbf{k} \delta
\rho^f_\mathbf{q-k}$. It is the same as in a bulk system, with
connected quantities simply replacing the fluid structure factor and
direct correlation function.  This is true for $\Omega_{q}^2$, given
by Eq.~\eqref{freqmic}, as well. Thus, we find that the relaxing part
of the density fluctuations in a QA system just behaves dynamically
like the corresponding degree of freedom in the bulk. This is not
really surprising, since a similar correspondence is already visible
in the statics, for instance in the OZ equation \eqref{OZconn} or in
the convolution approximation \eqref{convconn}, which are both used in
the calculation of $V^{(2)}_{\mathbf{q},\mathbf{k}}$. Such a
correspondence was missing in all our attempts to derive a MCT
starting from the total density fluctuations and this is one of the
reasons for which the resulting equations were considered
unsatisfactory. $V^{(1)}_{\mathbf{q},\mathbf{k}}$ is less obvious and
combines features of $V^{(2)}_{\mathbf{q},\mathbf{k}}$ and of the
vertex for the tagged particle dynamics in the bulk
\cite{LesHouches}. In particular, like the latter, it diverges when
$q\to0$.

Finally, the above equations differ slightly from those reported in
Ref.~\cite{Kra05PRL}, where a first account of the theory was
given. There, $S^{b}_{|\mathbf{q-k}|}$ in Eq.~\eqref{vone} is replaced
by $S^{b}_{|\mathbf{q-k}|} - n_f \hat{c}^{b}_{|\mathbf{q-k}|}
(S^{c}_{|\mathbf{q-k}|})^2$, where $\hat{c}^{b}_q$ is the Fourier
transform of the blocked direct correlation function.  The only
difference between the two approaches is that $\delta
\rho^f_\mathbf{k} \langle \rho^f_\mathbf{q-k} \rangle$ was not
included as a slow variable in the earlier mode-coupling scheme. A
priori, this is an unjustified approximation. Indeed, as shown in
Appendix \ref{app.derivation}, the coupling to this variable is
actually so strong, that, when both $\delta \rho^f_\mathbf{k}
\rho^m_\mathbf{q-k}$ and $\delta \rho^f_\mathbf{k} \langle
\rho^f_\mathbf{q-k} \rangle$ are considered, the contribution of the
former becomes identically zero, while the effect of the random
environment is integrally transferred to the latter.  There are
nevertheless circumstances in which the absence of $\delta
\rho^f_\mathbf{k} \langle \rho^f_\mathbf{q-k} \rangle$ looks perfectly
well motivated, for instance when a single particle is moving in the
porous matrix, corresponding to the limit $n_f\to0$, or when the
adsorbed fluid is an ideal gas \cite{IdealGas}. In both cases, the
present MCT reduces to $\Omega_{q}^2=q^2 k_B T/m$ and
\begin{equation} \label{lorentz}
m^{(\text{MC})}_q(t) = \int \frac{d^3\mathbf{k}}{(2\pi)^3} \left[
\frac{\mathbf{q}\cdot(\mathbf{q-k})}{q^2} \right]^2
\hat{h}^{b}_{|\mathbf{q-k}|} \phi_{k}(t),
\end{equation}
where $\hat{h}^{b}_q$ is the Fourier transform of the blocked total
pair correlation function. But, since the only forces exerted on the
fluid are those due to the random matrix, on physical grounds, the
only mode-coupling contributions to the relaxation kernel are expected
to come from $\delta \rho^f_\mathbf{k} \rho^m_\mathbf{q-k}$. All
previous theories of the Lorentz gas have been based on this insight,
which leads to the same equations as above, except that
$\hat{h}^{b}_{|\mathbf{q-k}|}$ in Eq.~\eqref{lorentz} is replaced with
$\hat{h}^{b}_{|\mathbf{q-k}|} - \hat{c}^{b}_{|\mathbf{q-k}|}$
\cite{GotLeuYip81PRA,Leu83PRA,Sza04EL}. Note that, for an adsorbed
ideal gas, both theories correctly predict that the dynamics is
independent of $n_f$.
 
It seems thus that there is a subtle interplay between the
approximations involved in the derivation of the MCT and the peculiar
structure of the static correlations in the QA mixture. At present, it
is not clear what to conclude from this observation, but it is
interesting that the difference between the two MCT schemes involves
the blocked direct correlation function \cite{NoteBDCF}. Indeed, this
is a rather delicate object, not easily captured by simple
approximations. This is best understood in the replica framework,
where $c^{b}(r)$ is obtained as the zero replica limit of the direct
correlation function between two non-interacting fluid replicas only
correlated through their common interaction with the matrix
\cite{GivSte92JCP,LomGivSteWeiLev93PRE,GivSte94PA,RosTarSte94JCP}.  It
results that $c^{b}(r)$ has a highly non-additive character and that,
for instance, most simple closures of the replica OZ equations fail to
provide expressions for this function which do not vanish
identically. Thus, it is a rather nontrivial finding that the MCT
approximation scheme is actually sensitive to the existence of this
function.

\section{Self-induced glassiness versus quenched random environment}

From the previous discussion, it is clear that the separation of the
fluid density fluctuations into relaxing and frozen parts plays a
crucial role in the derivation of the MCT for QA systems. In that
case, the freezing is of static origin, due to the random external
field generated by the quenched matrix.  But, as it is well known from
the MCT for bulk fluids, freezing of the density fluctuations may also
occur dynamically, when the system enters in the ideal glassy
state. It seems thus interesting to compare both situations. This can
be seen as a case of self-induced versus quenched disorder, similar to
what has been discussed many times in the literature
\cite{BouMez94JP1,MarParRit94JPA1,CugKurParRit95PRL,CugKurMonPar96JPA}.

The residual relaxation of a bulk fluid in its ideal glassy state has
been studied in Ref.~\cite{GotMay00PRE}. It is described by
mode-coupling equations of the same form as above, except that the
characteristic frequency is given by
\begin{equation} \label{freqmicresidual}
\Omega_{q}^2=\frac{q^2 k_B T}{m \left\{(1-f_q) S_q \right\}},
\end{equation}
and the vertices are
\begin{widetext}
\begin{subequations} \label{verticesresidual}
\begin{equation} \label{vtworesidual}
V^{(2)}_{\mathbf{q},\mathbf{k}} = \frac{1}{2} n_f \left\{(1-f_q) S_q
\right\} \left[\frac{\mathbf{q}\cdot\mathbf{k}}{q^2} \hat{c}_k +
\frac{\mathbf{q}\cdot(\mathbf{q-k})}{q^2}
\hat{c}_{|\mathbf{q-k}|}\right]^2 \left\{(1-f_k) S_k\right\}
\left\{(1-f_{|\mathbf{q-k}|}) S_{|\mathbf{q-k}|}\right\}
\end{equation}
and
\begin{equation} \label{voneresidual}
V^{(1)}_{\mathbf{q},\mathbf{k}} = n_f \left\{(1-f_q) S_q \right\}
\left[ \frac{\mathbf{q}\cdot\mathbf{k}}{q^2} \hat{c}_k +
\frac{\mathbf{q}\cdot(\mathbf{q-k})}{q^2} \hat{c}_{|\mathbf{q-k}|}
\right]^2 \left\{(1-f_k) S_k \right\} \left\{ f_{|\mathbf{q-k}|}
S_{|\mathbf{q-k}|} \right\},
\end{equation}
\end{subequations}
\end{widetext}
where $n_f$ is the density of the fluid, $S_q$ its structure factor,
$\hat{c}_q$ the Fourier transform of its direct correlation function,
and $f_q$ the Debye-Waller factor of the glass. Accordingly, $f_q S_q$
corresponds to the dynamically frozen part of the density
fluctuations, while $(1-f_q) S_q$ corresponds to their relaxing part.

The analogies between these equations and Eqs.~\eqref{freqmic} and
\eqref{vertices} are striking. There is almost a perfect
correspondence between $(1-f_q) S_q$ and $S^c_q$ on the one hand, and
$f_q S_q$ and $S^b_q$ on the other hand. Thus, we find that,
irrespective of the mechanism of freezing of the fluid density
fluctuations, the resulting relaxing and frozen contributions
essentially play the same role in both systems.

But there remains one significant difference between both sets of
equations. Indeed, one of the Fourier transformed direct correlation
functions in Eq.~\eqref{voneresidual}, the one which carries the same
wavevector as the frozen part of the structure factor, is replaced by
a simple constant factor $1/n_f$ in Eq.~\eqref{vone}. The origin of
this constant term should probably be traced back to the asymmetric
nature of the QA system, where the fluid reacts to the matrix but not
the other way around, resulting in a non-equilibrium character of its
static and dynamical correlations, in the sense that the matrix is not
equilibrated with the fluid. It would then reflect the lack of
dynamical self-consistency between the glassy dynamics in a QA system
and the frozen background on top of which it develops.
 
It is tempting to try and obtain the MCT for the QA binary mixture as
a limiting case of the MCT for the ordinary binary mixture
\cite{Got87NATO}, where one component would become the quenched
matrix. The above discussion shows that this is not possible. Indeed,
based on heuristic considerations, one can get close to
Eqs.~\eqref{freqmicresidual} and \eqref{verticesresidual}, for
instance by canceling all terms which would result in a time
dependence of the fluid-matrix and matrix-matrix correlations, but the
non-equilibrium density factor identified above seems impossible to
generate from the theory for the fully annealed system. Thus, as far
as the development of theoretical approaches is concerned, and this
statement is probably not restricted to the MCT framework, the
dynamics of a QA system should definitely be considered from the start
as different from an infinite mass (for Newtonian systems) or a zero
bare diffusivity (for Brownian dynamics) limit of the dynamics of a
fully annealed mixture. This is actually not so unexpected since,
already in the statics, similar difficulties were met in early
attempts to derive the equations valid for the QA mixture starting
from those describing fully equilibrated systems (see the discussion
of Refs.~\cite{MadGla88JSP,Mad92JCP,Cha91JPCM} in
Ref.~\cite{GivSte94PA}).

Thus, the present theory is not a special case of earlier
mode-coupling studies of the dynamics of particles moving in a glassy
matrix \cite{BosTha87PRL}. There, the MCT for binary mixtures was used
in a regime where one component, made of big particles, was completely
glassy, i.e., both the collective and tagged particle correlators did
not relax to zero, while the other component, made of small spheres,
was not necessarily localized, i.e., the tagged particle correlators
could relax to zero. In fact, both theories are complementary. In the
early approach, the glassy matrix has to be the product of the
self-consistent mode-coupling dynamics, but the model incorporates
thermal fluctuations of the solid and aspects of its response to the
presence of the fluid, while in the present approach, one has much
more freedom to choose the structure of the confining medium,
including realistic models of porous solids \cite{KraKieRosTar01JCP},
but the disordered matrix is rigorously inert. There are anyway
qualitative analogies between both models. For instance, there is a
clear link between the fact that the collective motion of the small
particles in a binary mixture necessarily becomes non ergodic at the
same point as that of the big particles and the unavoidable existence
of static blocked correlations for a fluid in a random matrix.

\section{Dynamical phase diagrams}

We now move to the quantitative predictions of the theory, which
require numerical solutions of the MCT equations.  In this section, we
report dynamical phase diagrams, obtained by mapping, in the parameter
space of a given model, the domain where $\lim_{t\to\infty}
\phi_q(t)=0$, corresponding to the ergodic fluid phase, and the one
where $\lim_{t\to\infty} \phi_q(t)=f_q\neq0$, corresponding to the
non-ergodic ideal glassy state. The interface between the two domains
forms the ideal liquid-glass transition manifold; in the present work,
it will always be a line, since only systems with a two-dimensional
parameter space are considered. Details of the dynamical changes when
crossing this line are discussed in the next section.

As already mentioned in Sec. II, a significant difference between
the MCT equations for QA systems and those for bulk glassformers is
the presence of a linear term in the memory kernel \eqref{kernel}. It
results that the discontinuous or type B transition scenario known
from the bulk, where the infinite time limit of $\phi_q(t)$ jumps
discontinuously from zero to a finite value when going from the liquid
to the glass, is not the only possibility anymore. Continuous or type
A transition scenarios, where $f_q$ grows continuously from zero when
entering the glassy phase, indeed become possible.

This is best understood by reference to the so-called $F_{1n}$
($n\ge2$) schematic models \cite{LesHouches}, in which the time
evolution of a single correlation function $\phi(t)$ is ruled by a two
parameter memory kernel of the form
\begin{equation}
m_{1n}(t)=v_1 \phi(t) + v_n \phi(t)^n, \quad v_1,v_n\ge0.
\end{equation}
It can be readily shown that, when the second term dominates ($v_1$
small), these models have a line of type B transitions starting at
$v_1=0,v_n=n^n/(n-1)^{n-1}$, and, when the first term dominates ($v_n$
small), they have a line of type A transitions starting at
$v_1=1,v_n=0$.  The study of the $F_{1n}$ models also gives us
information on the possible topologies of the dynamical phase
diagrams, which depend on the way the transition lines meet. Two
simple cases are found \cite{LesHouches}. If $n=2$, the two lines join
smoothly at a common endpoint, where a topologically stable degenerate
$A_3$ singularity is located. If $n>2$, the two lines intersect and,
in the glassy domain, only the extension of the type B transition line
subsists beyond the intersection, forming a glass-glass transition
line terminated by an ordinary $A_3$ singularity.

It turns out that these two prototypical shapes of phase diagrams are
obtained with two very simple, closely related QA mixture models, to
which the present study is restricted.  In both, the fluid-fluid and
fluid-matrix interactions are pure hard core repulsions of the same
diameter $d$. The only difference lies in the matrix correlations. In
model I, the matrix configurations are assumed to be quenched from an
equilibrium fluid of hard spheres of diameter $d$, so that the matrix
particles do not overlap, while in model II the matrix particles are
completely uncorrelated and overlap freely.  In the following, both
systems will be parametrized by the two dimensionless densities
$\phi_f=\pi n_f d^3/6$ and $\phi_m=\pi n_m d^3/6$, and the
Percus-Yevick (PY) approximation
\cite{GivSte92JCP,LomGivSteWeiLev93PRE,GivSte94PA,MerLevWei96JCP} will
be used to compute the required structural quantities. Note that, in
this approximation, $c^b(r)\equiv0$, so that the difficulties
mentioned at the end of Sec.~II are irrelevant.

The non-ergodicity parameter $f_q$ is a solution of the nonlinear set
of equations
\begin{equation}\label{nonlinear}
\frac{f_q}{1-f_q}=\int \frac{d^3\mathbf{k}}{(2\pi)^3} \left[
V^{(2)}_{\mathbf{q},\mathbf{k}} f_{k} f_{|\mathbf{q-k}|} +
V^{(1)}_{\mathbf{q},\mathbf{k}} f_{k} \right], 
\end{equation}
which has to be solved numerically in order to locate the liquid and
glassy phases when $\phi_f$ and $\phi_m$ are varied. All computations
in the present work have been achieved using the method of
Ref.~\cite{FraFucGotMaySin97PRE}, to which the interested reader is
referred for technical details, and we only provide the quantitative
information needed to reproduce the present results, i.e., that the
wavevector integrals have been discretized to points on a grid of
$128$ equally spaced values with step size $\Delta\simeq0.3835/d$,
starting at $q_\text{min}=\Delta/2$ \cite{notePY}. We have checked by
two means that this discretization is not too coarse, in particular
with respect to the small $q$ divergence of the memory kernel which is
cut off. First, test calculations at various fluid and matrix
densities have been performed on a finer $q$ grid, hence with a
smaller $q_\text{min}$. Second, in the $n_f\to0$ limit, we have
compared our prediction for the diffusion-localization point of model
I with Leutheusser's analytic result obtained within an additional
hydrodynamic approximation which allows to integrate exactly over the
full $q$ range \cite{Leu83PRA}.  In all cases, only modest
quantitative differences of a few percents on the location of the
transition points were found.

\begin{figure*}
\includegraphics*{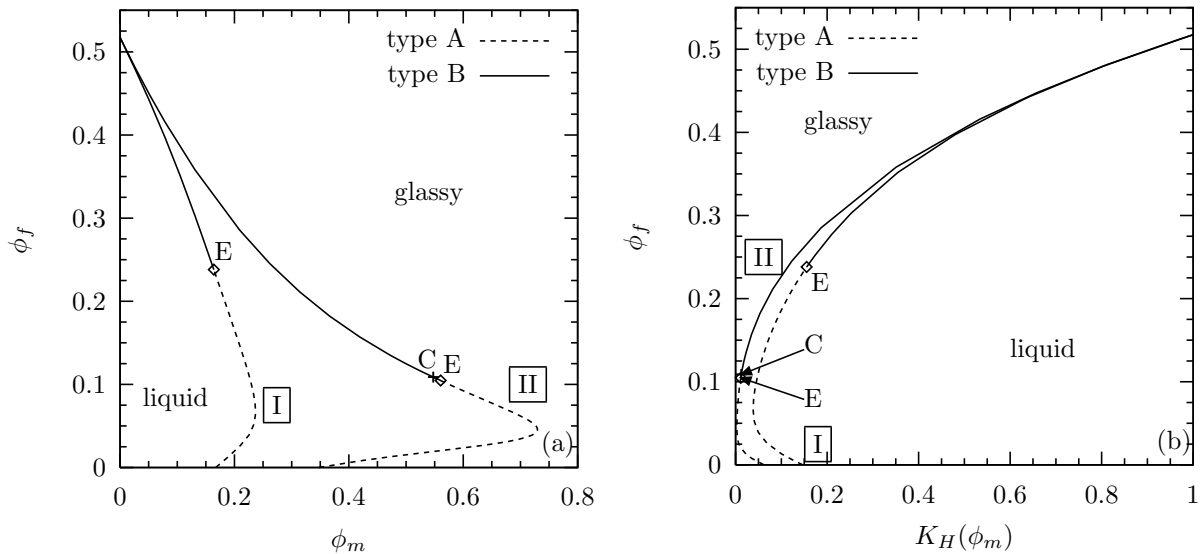}
\caption{\label{figtrans} Dynamical phase diagrams of two types of
  hard sphere quenched-annealed binary mixtures. (a) In the
  $(\phi_m,\phi_f)$ plane. (b) In the $(K_H(\phi_m),\phi_f)$
  plane. Model I: non-overlapping matrix particles. Model II: freely
  overlapping matrix particles. The $A_3$ singularities are indicated
  with diamonds and denoted by E, while the crossing point between the
  type A and B transition lines in model II is indicated by a cross
  and denoted by C. }
\end{figure*}

When dealing with complex transition scenarios, including higher-order
singularities and glass-glass transition lines, a useful quantity to
consider is the largest eigenvalue $E$ of the stability matrix of the
set of equations \eqref{nonlinear}. Indeed, there holds $E\leq1$, and
$E$ goes to $1$ when a transition is approached from the strong
coupling side \cite{GotSjo95JMAA}. Thus, $1-E$ can be used as a
convergence criterion for the determination of the transition points
\cite{Spe04PRE}. In this work, $1-E\leq10^{-3}$ was assured for
ordinary transition points and $1-E \leq 10^{-5}$ was required in the
regions where the transition lines of types A and B meet.

The dynamical phase diagrams of models I and II are reported in
Fig.~\ref{figtrans}.  In the left panel, they are plotted in the
$(\phi_m,\phi_f)$ plane. This is the obvious parameter space of the
problem, but, with this choice of variables, no account is given of
the differences in structure between the two matrix models. This
clearly limits the possibilities of a meaningful comparison between
the two systems. Thus, in the right panel, both QA mixtures have been
tentatively parametrized by the same physical constant, their Henry
constant $K_H(\phi_m)$. For systems with hard core interactions,
$K_H(\phi_m)$ is equal to the fraction of the total volume accessible
to the center of an adsorbate particle in a matrix of density $\phi_m$
and is probably the most simple and generic scalar quantity
characterizing the confining effect of a solid matrix on an adsorbed
fluid.  It is given by $\exp(-8\phi_m)$ for model II
\cite{KamMon91JCP}, and by $\exp[-\beta\mu_\text{ex}(\phi_m)]$ for
model I, where $\mu_\text{ex}(\phi_m)$ is the excess chemical
potential of the equilibrium hard sphere fluid with volume fraction
$\phi_m$ \cite{KieRosTarMon97JCP}. The latter has been estimated
following the compressibility route within the PY approximation
\cite{simpleliquids}.

As it can be readily seen in Fig.~\ref{figtrans}, both phase diagrams
essentially have the same overall shape, especially when they are
plotted as functions of $K_H(\phi_m)$. Their main qualitative
difference at this global scale, which is the difference in concavity
of their upper parts visible in Fig.~\ref{figtrans}(a), turns out to
be rather insignificant, as it is the simple and direct consequence of
the fact that the strength of confinement increases more slowly with
$\phi_m$ in model II than in model I, because of the overlapping
matrix particles.  It completely disappears in Fig.~\ref{figtrans}(b).

The nature of the ideal glass transitions met in the phase diagrams is
just as expected from the study of schematic models. Starting from the
type B liquid-glass transition point for the bulk (at $\phi_m=0$),
where the memory kernel is purely quadratic, a line of type B
transitions develops when increasing $\phi_m$, and, starting from the
type A diffusion-localization point (at $\phi_f=0$), where the memory
kernel is purely linear, a line of type A transitions emerges when
increasing $\phi_f$.  The same is true of the way the two lines meet
in the phase diagrams. On the one hand, for model I, the type A and B
lines have a common endpoint, denoted by E, where they smoothly join
and form a degenerate $A_3$ singularity, like in the $F_{12}$
model. On the other hand, for model II, the two lines intersect at a
crossing point C and the extension of the type B transition line in
the glassy domain becomes a glass-glass transition line ending with an
ordinary $A_3$ singularity, denoted by E as well. This is the scenario
obtained with the $F_{1n}$ models, for $n>2$. Note that, to our
knowledge, this is the first time that these widely studied one
equation toy models find physical realizations as fluid systems.

The glass-glass transition line of model II, located between points C
and E, is barely visible at the scale of Fig.~\ref{figtrans}. It is
short and not well separated from the type A liquid-glass transition
line. Taking into account the idealized nature of the predictions of
the MCT, it would probably be impossible to detect it unambiguously in
computer simulations of this system. Only the very specific features
of the dynamics in the vicinity of a higher-order singularity, be it
degenerate or not, should be visible, like logarithmic decay laws and
subdiffusive behaviors
\cite{LesHouches,DawFofFucGotSciSpeTarVoiZac01PRE,Spe04PRE,GotSpe02PRE}.
But, at least, the present calculation shows that this scenario can
actually be realized and that this does not require any exotic
physical ingredient. Now that a suitable starting point is available,
by playing with the parameters of the model, one can try and obtain a
system for which this glass-glass transition line would be extended
enough for its signatures to be observable in simulations.

\begin{figure}
\includegraphics*{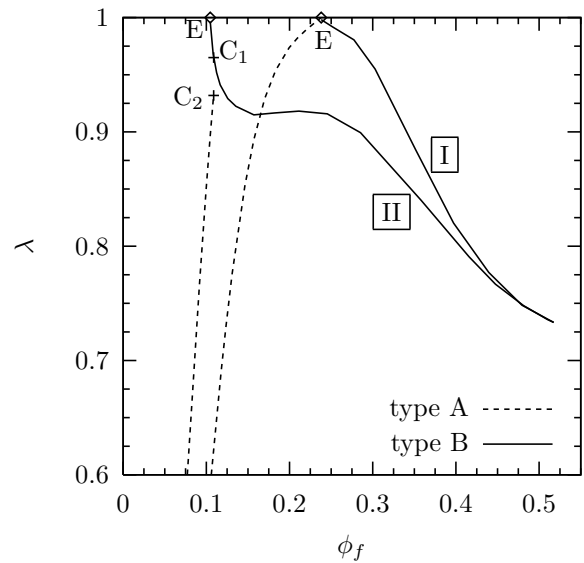}
\caption{\label{figlambda} Exponent parameter $\lambda$ along the
  transition lines of models I and II. The lowest part, where
  $\lambda$ goes to zero with $\phi_f$, has been omitted for
  readability. For both models, E denotes the $A_3$ singularity, with
  $\lambda=1$. C$_1$ and C$_2$ delimit the discontinuity associated
  with the crossing point between the type A and B transition lines in
  model II.}
\end{figure}

A final requirement for a complete characterization of the transitions
studied in this work is the knowledge of the so-called exponent
parameter $\lambda$ \cite{LesHouches}. It determines many aspects of
the dynamics near a transition (see the next section) and, for this
reason, plays a crucial role in the theory.  One has $0\le\lambda\le1$
and $1/2\le\lambda\le1$ for type A and B transitions,
respectively. Also, $\lambda$ reaches $1$, its maximum value, at
endpoint singularities and jumps discontinuously at crossing
points. This is thus a useful parameter to follow during the
computation of the phase diagrams. It is plotted in
Fig.~\ref{figlambda} as a function of $\phi_f$ at the transition. For
model II, points C$_1$ and C$_2$ mark the discontinuity associated
with the crossing point C in Fig.~\ref{figtrans} and the existence of
glass-glass transitions is clearly visible, with values of $\lambda$
given by the line between points C$_1$ and E. For the same model, a
nonmonotonic variation of $\lambda$ along the type B line can be noted
as well.

In addition to these transition scenarios, and formally not related to
them, another remarkable prediction of the present theory is a reentry
phenomenon for matrix densities higher than the localization threshold
(obtained for $\phi_f=0$). Indeed, as shown in Fig.~\ref{figtrans}(a),
for a given, not too high $\phi_m$ in this domain, ergodicity can be
broken either by an increase or a decrease of the fluid density.

In hard core systems, freezing by an increase of the fluid density can
be qualitatively understood from simple free volume arguments, which
provide a direct explanation for the decrease of $\phi_f$ at the
transition as a function of $\phi_m$ in the upper part of the phase
diagrams. Because of the volume excluded by the matrix particles, the
larger the matrix density is, the smaller the fluid density has to be
for structural arrest to occur.  From the results for model I, we
might further note that this decrease is such that the total compacity
$\phi_f+\phi_m$ at the transition is a decreasing function of $\phi_m$
as well, reflecting the fact that the inclusion of immobile matrix
particles in the system slows down the dynamics more efficiently than
the inclusion of the same amount of mobile fluid particles. Such a
behavior, which is hardly surprising, has already been observed in
molecular dynamics simulations \cite{Kim03EL,ChaJagYet04PRE}.

The possibility of an ergodicity breaking transition by a decrease of
the fluid density, which is reflected in the bottom part of the phase
diagrams by the increase with $\phi_m$ of the transition $\phi_f$, is
more unexpected. We interpret this prediction as the signature of a
delocalization phenomenon induced by fluid-fluid interactions. More
precisely, we propose that the occasional collisions between the fluid
particles at low $\phi_f$ can destroy the dynamical correlations
responsible for the localization of individual particles in dense
enough matrices.  For this process to provide an efficient ergodicity
restoring relaxation channel, a reasonable criterion is that the
localization domains should overlap to allow the fluid particles to
interact. Thus, the localization length computed at $\phi_f=0$, which
decreases when $\phi_m$ increases, should be comparable to the average
distance between two fluid particles, which decreases when $\phi_f$
increases. It is then immediate that, starting in the localized state,
the larger $\phi_m$ is, the higher $\phi_f$ has to be in order to
restore ergodicity. The physical implications of this result will be
discussed in more details in the last section.

Dynamical scenarios involving reentrant glass transition lines,
higher-order singularities, and glass-glass transition lines, have
already been found for colloidal suspensions with short-ranged
attractions \cite{DawFofFucGotSciSpeTarVoiZac01PRE}. By analyzing
various contributions to the memory kernel, these features were shown
to result from the interplay of two well defined phenomena, cage
effect and bond formation, driven by the hard core and attractive
parts of the interaction, respectively. It seems thus interesting to
attempt such an analysis for the present problem.  

Guided by the results of Sec.~III, we propose that a contribution
represented by a memory kernel $m^{(\text{cage})}_q(t)$ derived from
$m^{(\text{MC})}_q(t)$ by replacing $1/n_f$ in Eq.~\eqref{vone} with
$\hat{c}^{c}_{|\mathbf{q-k}|}$ should be isolated. Indeed, the
resulting expression then coincides with the one describing the
residual dynamics of a bulk ideal glass. Therefore, one can reasonably
expect that, for a QA system, $m^{(\text{cage})}_q(t)$ will provide a
fair representation of the mechanism of caging by fluid particles in
the presence of permanent density fluctuations, which is precisely the
one at work in bulk glassy systems.  The remaining linear kernel
$m^{(\text{conf})}_q(t) = m^{(\text{MC})}_q(t) -
m^{(\text{cage})}_q(t)$ can then be attributed to the
confinement-specific phenomena, i.e., the localization effect of the
matrix combined with the unusual decorrelation mechanism due to
fluid-fluid collisions discussed above. In principle, one could try to
separate these two processes, using the fact that, for an ideal gas,
only the localization effect is present and leads to a matrix density
at the transition which is independent of the fluid density.  However,
already with the present limited separation, the somewhat artificial
character of the procedure shows up in the form of negative values of
$m^{(\text{conf})}_q(t)$, which restrict the density domain where
stable solutions of the MCT equations can be found, so no further
decomposition of $m^{(\text{conf})}_q(t)$ was attempted.

\begin{figure}
\includegraphics*{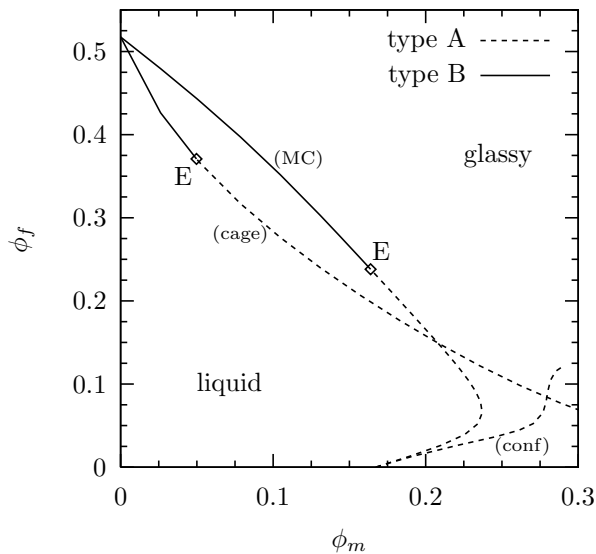}
\caption{\label{figfictitious} Real and hypothetical dynamical phase
  diagrams of model I, computed with the total and partial memory
  kernels $m^{(\text{MC})}_q(t)$, $m^{(\text{cage})}_q(t)$, and
  $m^{(\text{conf})}_q(t)$. The curves are labeled accordingly and the
  degenerate $A_3$ singularities are denoted by E.}
\end{figure}

The hypothetical phase diagrams computed with the partial kernels
$m^{(\text{cage})}_q(t)$ and $m^{(\text{conf})}_q(t)$ for model I are
reported in Fig.~\ref{figfictitious}, where they are compared to the
one obtained with the full $m^{(\text{MC})}_q(t)$. Clearly, the
reentry phenomenon observed in the complete phase diagram can be
explained by the interplay of the two contributions discussed
above. But, at variance with the colloidal systems, the higher-order
singularity is not located in the domain where they cross over. The
interpretation of this finding is ambiguous. On the one hand, since a
higher-order singularity is found on the transition line computed with
$m^{(\text{cage})}_q(t)$ as well, one might argue that the singularity
should be considered as an integral part of the scenario of caging by
fluid particles. Then, in the relevant domain, confinement would
simply appear as a modifier of the cage structure, progressively
changing the nature of the ideal jamming transition from discontinuous
to continuous.  On the other hand, the bifurcation analysis of the MCT
scenario shows that a higher-order singularity is necessarily formed
when two lines of type A and B transitions meet.  Then, since the type
A and B lines arise from points representative of systems ruled by
confinement and bulk caging, respectively, one might consider that the
singularity is the product of the interplay of these two phenomena.
In favor of the latter interpretation, it has been recently suggested
that higher-order singularities generically result from a competition
between different arrest mechanisms \cite{MorCol06JCP}.  Note however
that the conclusions of Ref.~\cite{MorCol06JCP} are drawn from the
consideration of bulk systems only, which necessarily enter into the
glassy state through type B transitions and for which there is no
mathematical constraint in the theory imposing a priori the existence
of a singularity. It is not immediate that the proposed statement is
valid or even needed for systems where type A and B liquid-glass
transitions coexist.

\section{Typical transition scenarios}

\begin{figure*}
\includegraphics*{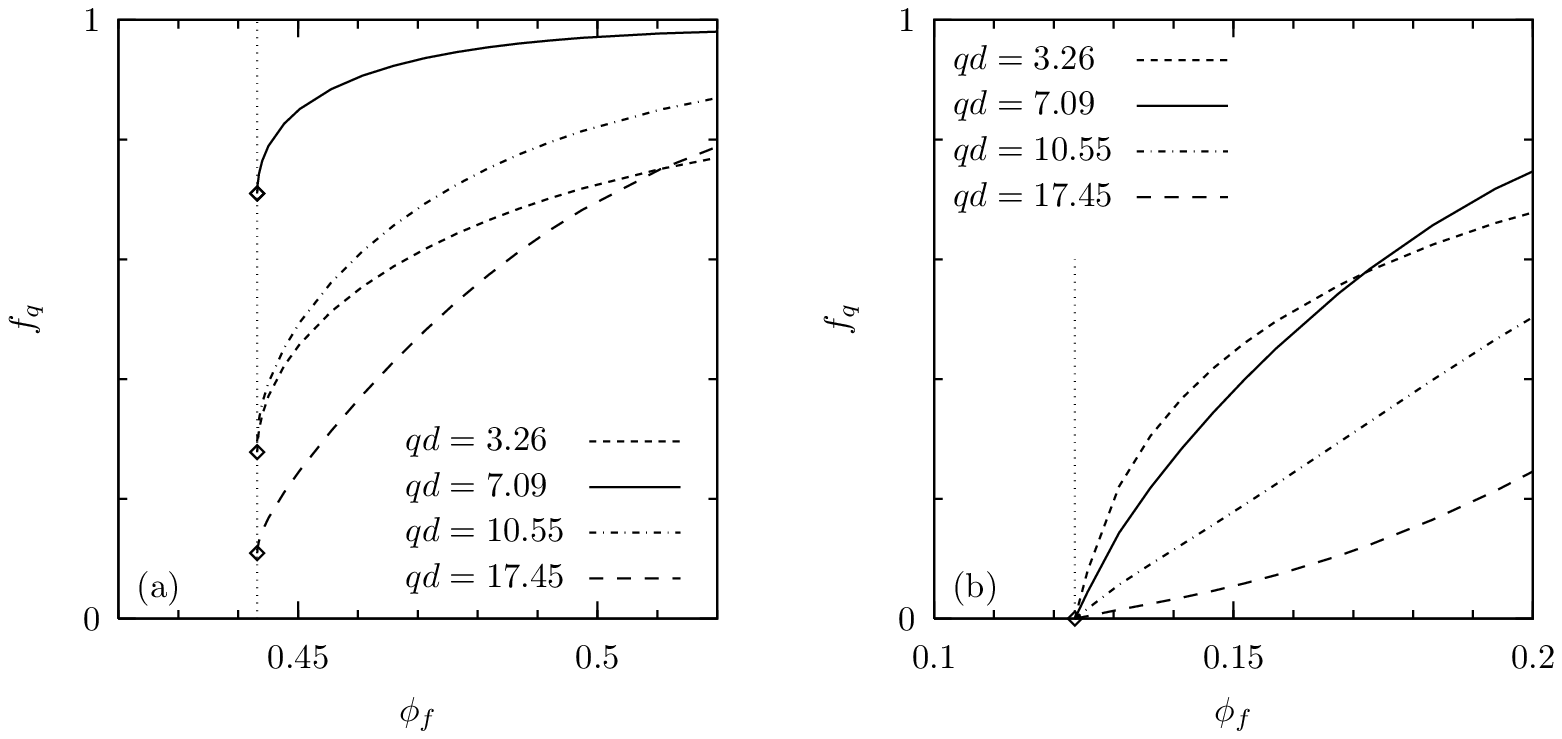}
\caption{\label{fignonergvsphif} Fluid density dependence of the
  non-ergodicity parameter $f_q=\lim_{t\to\infty} \phi_q(t)$ for model
  I at two matrix densities. (a) Type B transition for $\phi_m=0.05$.
  (b) Type A transition for $\phi_m=0.22$. Data for four
  representative wavevectors are shown. Vertical dotted lines and
  diamonds indicate the ideal glass transition.  }
\end{figure*}

It results from the previous section that, if one stays away from the
higher-order singularities, the crossing points, and the glass-glass
transitions, which require fine tuning of the parameters of the
models, the present MCT for QA mixtures predicts two generic
liquid-glass transition scenarios. One is discontinuous or type B, the
other is continuous or type A.  In this section, we discuss the
different features of these transitions which are relevant for
comparisons of experimental or simulation data with the predictions of
the theory. The analytic results will be quoted without their proofs,
which can be found in Ref.~\cite{LesHouches}. They will be illustrated
with detailed computations for model I at two matrix densities,
$\phi_m=0.05$ and $\phi_m=0.22$. For the former value, a type B
transition occurs at $\phi^c_f\simeq 0.443$, with $\lambda \simeq
0.774$. For the latter, a type A transition is found at
$\phi^c_f\simeq 0.124$, with $\lambda \simeq 0.729$. For this matrix
density, we consider the type A transition on the upper branch of the
phase diagram, as it is of greater relevance for the problem of the
glass transition in confinement. In both cases, keeping $\lambda\leq
0.8$ allows one to consider that the transitions are far enough from
the higher-order singularity.

Here, a comment on the method of solution of the mode-coupling
equations is in order. Indeed, as mentioned in the previous section,
it involves a cutoff of the low $q$ divergence of the memory
kernel. Since it has been demonstrated that this divergence can change
the qualitative properties of the solutions in the asymptotic regime
near the transition \cite{Got81PMB,Leu83PRAb,HofFreFraUNPUB}, this
approximation and the use of the results of Ref.~\cite{LesHouches},
which have been obtained under the assumption of nonsingular vertices,
might look problematic. In fact, this is not the case for physical
reasons. Indeed, in the Lorentz gas limit, the low $q$ singularity of
the memory function has been shown to be an ill feature of the
mode-coupling approximation \cite{Leu83PRAb}. Thus, from a physical
point of view, the discretized equations considered in the present
calculation are actually more satisfactory than the continuous ones
and the results reviewed in Ref.~\cite{LesHouches} can legitimately be
applied.

\begin{figure*}
\includegraphics*{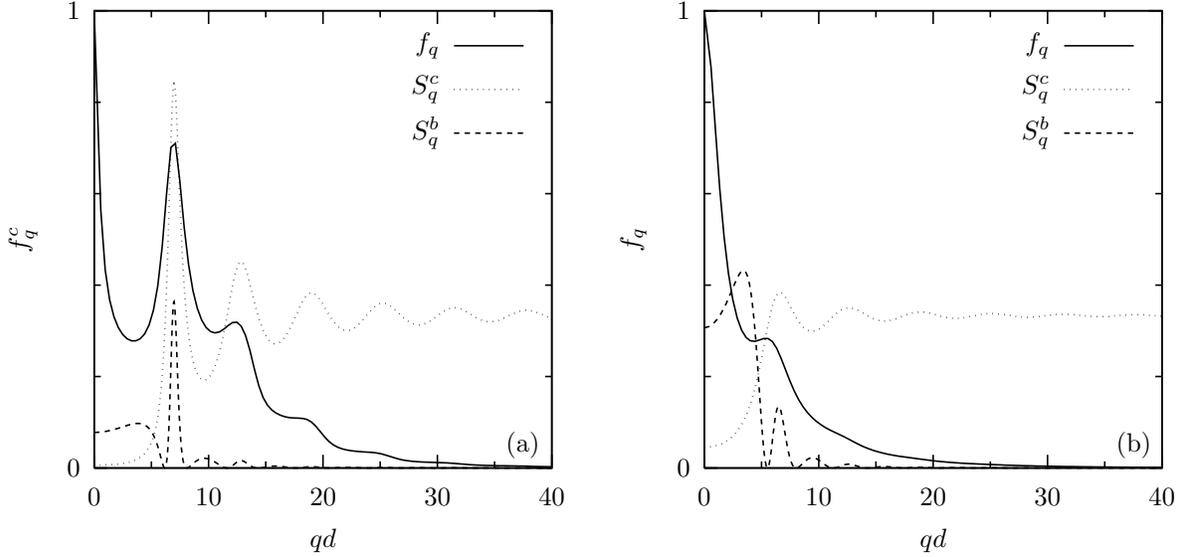}
\caption{\label{fignonergvsq} Wavevector dependence of the
  non-ergodicity parameter $f_q=\lim_{t\to\infty} \phi_q(t)$ for model
  I at two matrix and fluid densities. (a) At the type B transition
  for $\phi_m=0.05$. (b) Near the type A transition for $\phi_m=0.22$,
  at $\phi_f=1.1\phi_f^c$. In both panels, $S^c_q$ and $S^b_q$ at
  $\phi_f^c$ are reported for reference ($S^c_q$ has been divided by
  three for readability). }
\end{figure*}

We first consider the density dependence of the non-ergodicity
parameter $f_q$, shown in Fig.~\ref{fignonergvsphif}. As it should,
$f_q$ takes a finite value at the ideal glass transition point of type
B, while it grows continuously from zero at the type A transition. In
the glassy state, the bifurcation analysis of Eq.~\eqref{nonlinear} to
leading order yields two universal power law behaviors,
\begin{equation} \label{cusp}
f_q-f^c_q \propto (\phi_f-\phi^c_f)^{1/2}
\end{equation} 
for a type B scenario, and
\begin{equation} \label{slope}
f_q \propto (\phi_f-\phi^c_f)
\end{equation} 
near a type A transition.  Accordingly, in Fig.~\ref{fignonergvsphif},
the curves corresponding to the type B and A transitions start with
infinite and finite slopes, respectively.

The wavevector dependence of $f_q$, though not universal, is an
important prediction of the MCT as well. It is reported in
Fig.~\ref{fignonergvsq}. In Fig.~\ref{fignonergvsq}(a), $f_q$ is shown
at the type B transition, while in Fig.~\ref{fignonergvsq}(b), since
$f^c_q=0$ at a type A transition, the results at $\phi_f=1.1\phi_f^c$
are plotted. For reference, the two relevant structure factors $S^c_q$
and $S^b_q$ at $\phi_f^c$ are also given. Note that in both cases,
$S^b_q$ shows maxima both where $S^c_q$ has maxima or minima. At the
type B transition, except for the peak at $q=0$ where $f_q$ reaches 1
as a consequence of the diverging kernel, the non-ergodicity parameter
is very similar to the one found for a bulk hard sphere fluid and
oscillates with $S^c_q$, which precisely represents bulk-like
correlations. The overall amplitude is smaller than in the bulk,
reflecting the fact that, when $\phi_m$ increases, the system evolves
towards a continuous transition scenario. In comparison, the
non-ergodicity parameter near the type A transition appears rather
featureless. $f_q$ simply decreases from $1$ at $q=0$, with, as
$\phi_f$ is increased, a small shoulder developing in the wavevector
regime where $S^c_q$ has its main peak and $S^b_q$ its second
peak. With the present models where the fluid and matrix particles
have the same size, it is not clear which changes in the static
correlations are actually responsible for the growth of this
contribution.

Beside these results for the infinite time limit of the density
correlation functions, the full dynamics is of great interest as
well. For this computation, which uses the algorithm described in
Ref.~\cite{FucGotHofLat91JPCM}, we follow
Ref.~\cite{FraFucGotMaySin97PRE} and reduce the generalized Langevin
equation \eqref{eqlangevin} to its form valid for Brownian dynamics,
\begin{equation}
\tau_{q} \dot{\phi}_{q}(t) + \phi_{q}(t) + \int_0^t d\tau m_q(t-\tau)
\dot{\phi}_{q}(\tau)=0,
\end{equation}
with
\begin{equation}
\tau_{q}=t_\text{mic} S^{c}_q / (q d)^2
\end{equation}
and the initial condition $\phi_q(0)=1$. This simplification affects
the short time transient part of the dynamics, but not its long time
properties. In the following, the unit of time shall be chosen such
that $t_\text{mic}=160$.

\begin{figure*}
\includegraphics*{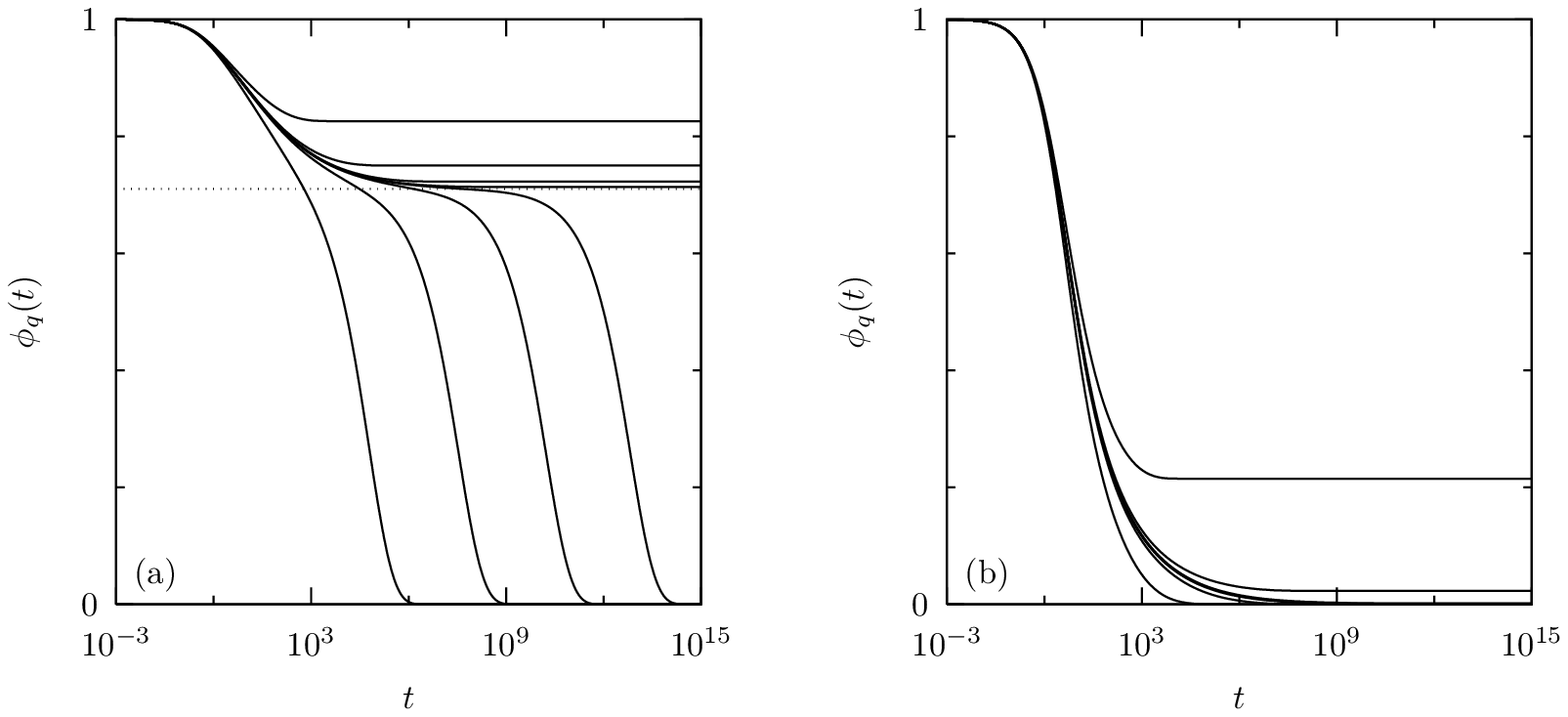}
\caption{\label{figcorrel} Time evolution of the connected density
correlation function $\phi_{q}(t)$ at $q \simeq 7.09/d$ for model I at
two matrix densities. (a) In the vicinity of the type B transition for
$\phi_m=0.05$; from left to right, bottom to top: $\phi_f =
0.99\phi_f^c$, $0.999\phi_f^c$, $0.9999\phi_f^c$, $0.99999\phi_f^c$,
$1.00001\phi_f^c$, $1.0001\phi_f^c$, $1.001\phi_f^c$, $1.01\phi_f^c$.
The horizontal dotted line marks $f^c_q$, the non-ergodicity parameter
at the transition. (b) In the vicinity of the type A transition for
$\phi_m=0.22$; from left to right, bottom to top: $\phi_f =
0.9\phi_f^c$, $0.99\phi_f^c$, $0.999\phi_f^c$, $0.9999\phi_f^c$,
$1.0001\phi_f^c$, $1.001\phi_f^c$, $1.01\phi_f^c$, $1.1\phi_f^c$.}
\end{figure*}

The time evolution of the density correlation function $\phi_{q}(t)$
at $q \simeq 7.09/d$, corresponding to the main peak of $S^c_q$, is
reported in Fig.~\ref{figcorrel} for state points in the vicinity of
the two transitions discussed above. The curves for other values of
$q$ are qualitatively similar. In Fig.~\ref{figcorrel}(a), the two
step dynamics typical of the discontinuous ideal glass transition
scenario and well known from the study of bulk systems is easily
recognized in the curves corresponding to the liquid state. The second
step, associated to the decay from the plateau where $\phi_{q}(t)
\simeq f^c_q$, obeys the so-called superposition principle, which
states that, in this regime and for a given $q$, the shape of the
dynamics is independent of the state point. The relaxation functions
only differ through the characteristic time scale $\tau_\alpha$, which
displays a power law divergence,
\begin{equation} \label{alpha}
\tau_\alpha \propto (\phi_f-\phi^c_f)^{-\gamma},
\end{equation}
when the transition is approached. One shows that
\begin{equation}
\gamma=\frac{1}{2a} + \frac{1}{2b},
\end{equation}
where the exponents $a$ and $b$ ($0<a<1/2$, $b>0$) are related to
$\lambda$ through
\begin{equation}\label{exponents}
\frac{\Gamma(1-a)^2}{\Gamma(1-2a)} =
\frac{\Gamma(1+b)^2}{\Gamma(1+2b)} = \lambda,
\end{equation}
$\Gamma$ denoting Euler's gamma function.  In the glassy state, only
the first relaxation step remains and, when $t$ goes to infinity,
$\phi_{q}(t)$ reaches $f_q$, which increases with $\phi_f$ as shown in
Fig.~\ref{fignonergvsphif}(a).

The dynamics near the type A transition visible in
Fig.~\ref{figcorrel}(b) looks significantly different, with a single
step relaxation scenario, both in the liquid and glassy phases. The
slowing-down of the dynamics manifests itself through a weak long time
tail which extends to longer times when $\phi_f$ is increased and
turns above $\phi_f^c$ into a finite asymptote which grows as shown in
Fig.~\ref{fignonergvsphif}(b).

There are nevertheless strong similarities between the two dynamics,
provided one concentrates on the time domain where $\phi_{q}(t) \simeq
f^c_q$ (i.e., where $\phi_{q}(t)$ is small for a type A
transition). For type B dynamics, this corresponds to the so-called
fast $\beta$ relaxation regime and we first specialize the discussion
to this case. Then, close enough to the transition, a reduction
theorem holds, according to which the wavevector and time dependencies
of $\phi_{q}(t)$ factorize, yielding
\begin{equation} \label{factor}
\phi_{q}(t) = f^c_q + h_q G(t).
\end{equation}
At the critical point, the critical decay law
\begin{equation} \label{critical}
G(t)=(t_0/t)^a
\end{equation}
is obtained, where $a$ is given by Eq.~\eqref{exponents} and $t_0$ is
a time scale obtained by matching the short and long time dynamics.
For finite values of $\phi_f-\phi^c_f$, one finds the scaling laws
\begin{equation} \label{scaling}
G(t) = c_\beta g_{\pm}(t/\tau_\beta)\qquad \phi_f-\phi^c_f\gtrless 0,
\end{equation}
where the master functions $g_{\pm}(\hat{t})$ are the solutions of 
\begin{equation}
\pm 1+\lambda g_{\pm}(\hat{t})^2 - \frac{d}{d \hat{t}} \int_0^{\hat{t}}
d\hat{\tau} g_{\pm}(\hat{t}-\hat{\tau}) g_{\pm}(\hat{\tau})=0,
\end{equation}
and the scaling variables obey
\begin{equation}
c_\beta \propto |\phi_f-\phi^c_f|^{1/2},\quad \tau_\beta/t_0 \propto
|\phi_f-\phi^c_f|^{-1/2a}.
\end{equation}
For small $\hat{t}$, we have
\begin{equation} \label{abovefc}
g_{\pm}(\hat{t}) = 1/ \hat{t}^a.
\end{equation}
For large $\hat{t}$, in the non-ergodic phase, $g_{+}(\hat{t})$ goes
to a constant and Eq.~\eqref{cusp} results from the expression of
$c_\beta$. In the same time regime, in the ergodic phase, another
power law behavior sets in for $g_{-}(\hat{t})$, yielding the
so-called von Schweidler decay law,
\begin{equation} \label{schweidler}
g_{-}(\hat{t}) = - B \hat{t}^b,
\end{equation}
where $b$ is given by Eq.~\eqref{exponents} and $B$ is a positive
constant which can be determined by matching Eq.~\eqref{schweidler}
with Eq.~\eqref{abovefc} for $\hat{t}\simeq 1$. By combining the power
law behaviors of $c_\beta$ and $\tau_\beta$ in the resulting
expression of $G(t)$, one recovers the divergence of $\tau_\alpha$,
Eq.~\eqref{alpha}.

Moving now to the type A transitions, one finds that, both in the
ergodic and non-ergodic states, $\phi_{q}(t)$ essentially behaves near
zero as it does when it approaches $f_q^c$ in the glassy phase in a
type B scenario. In particular, the critical decay law
\eqref{critical} remains valid at the transition. For finite values of
$\phi_f-\phi^c_f$ and independently of its sign, Eq.~\eqref{factor}
has to be modified in order to properly define $G(t)$ (see
Ref.~\cite{LesHouches} for details), then one finds that a scaling law
holds, of the form
\begin{equation}
G(t) = c g_{+}(t/\tau),
\end{equation}
where $g_{+}(\hat{t})$ is the same function as above. However, as
in Eq.~\eqref{slope}, the exponents characterizing the
scaling variables $c$ and $\tau$ are twice those for a type B
transition, i.e.,
\begin{equation} \label{scalA}
c \propto |\phi_f-\phi^c_f|,\quad \tau/t_0 \propto
|\phi_f-\phi^c_f|^{-1/a}.
\end{equation}

\begin{figure*}
\includegraphics*{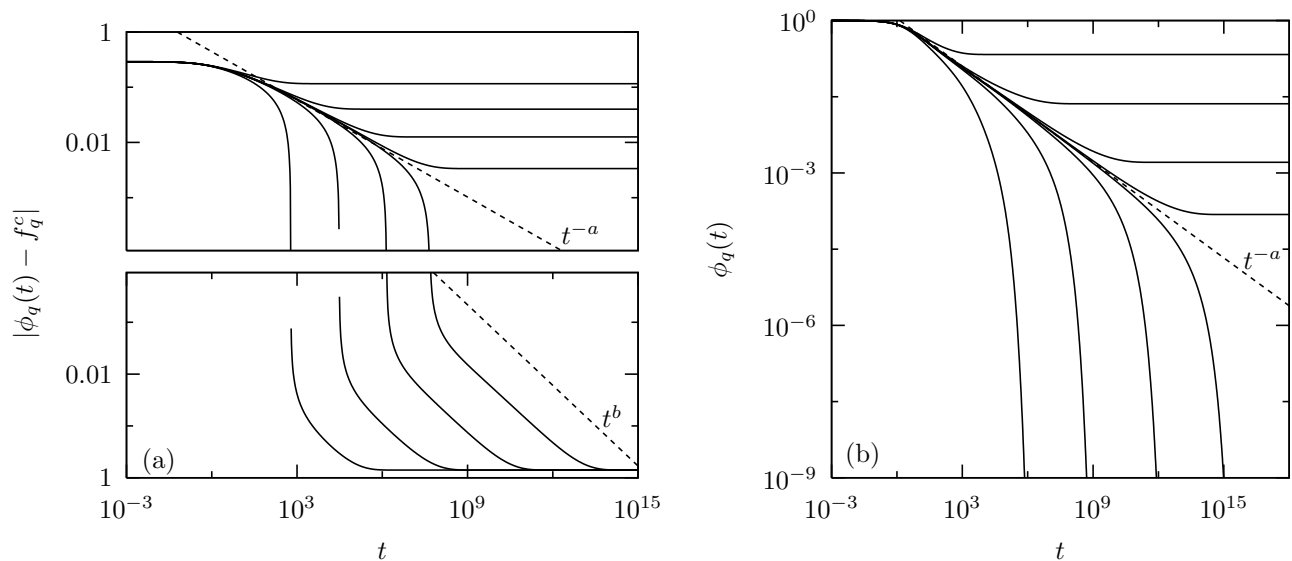}
\caption{\label{figcritical} Critical dynamics of the connected
density correlation function $\phi_{q}(t)$ at $q \simeq 7.09/d$ for
model I at two matrix densities.  (a) In the vicinity of the type B
transition for $\phi_m=0.05$. As it should, only the data for $\phi_f
< \phi_f^c$ appear in the bottom panel (note its reverted $y$ scale,
in order to make it graphically clear that, despite the use of the
absolute value, a decay law is represented). (b) In the vicinity of
the type A transition for $\phi_m=0.22$. See Fig.~\ref{figcorrel} for
attribution of the curves. The analytically derived power law
behaviors \eqref{abovefc} and \eqref{schweidler} are shown as dashed
straight lines.}
\end{figure*}

These behaviors are illustrated in Fig.~\ref{figcritical}, where the
time evolution of $|\phi_q(t)-f^c_q|$ (simply $\phi_q(t)$ for the type
A transition) at $q \simeq 7.09/d$ is plotted in a log-log scale in
order to evidence the power laws \eqref{abovefc} and
\eqref{schweidler}. In this graphs, an interesting consequence of the
scaling laws for small $\hat{t}$ is clearly visible, which is the
symmetric departure from the critical decay law \eqref{critical} at
long times for points located in the liquid and glassy phases at the
same distance from the transition. This is a particularly important
feature of the dynamics in the type A scenario.

The above results do not apply in the vicinity of higher-order
singularities or crossing points, where a refined mathematical
analysis is required \cite{LesHouches,Spe04PRE,GotSpe02PRE}. A
detailed discussion is beyond the scope of the present overview of the
theory and we shall only mention that such liquid-glass transition
points signal themselves in the dynamics through the appearance of
logarithmic decay laws
\cite{LesHouches,DawFofFucGotSciSpeTarVoiZac01PRE,Spe04PRE,GotSpe02PRE}.

Finally, already for bulk systems, it is often quite difficult to
unambiguously demonstrate that the above features are actually present
in some experimental or simulation data, since, as mentioned in the
introduction, the MCT offers an idealized picture of the glass
transition phenomenon and there are always significant alterations to
the theoretical scenario.  An additional difficulty can be anticipated
in the case of QA systems. Indeed, one usually has access to the total
density correlation function $\phi^T_q(t)$ and not to the connected
function $\phi_q(t)$. Both are related through
\begin{equation}
\phi^T_q(t)=\frac{S^{c}_q}{S^{ff}_q} \phi_q(t) +
\frac{S^{b}_q}{S^{ff}_q}.
\end{equation}
Thus, the glassy dynamics is modulated by static factors and in
particular develops itself on top of a state dependent static
background. The separation in the long time behavior of $\phi^T_q(t)$
of the evolutions which are of purely static origin from those which
characterize the glassy dynamics might then be delicate. This could be
especially critical for type A transitions, where the signatures of
the continuous transition have to be identified just on top of the
slowly drifting static contribution. Figures demonstrating the problem
can be found in Ref.~\cite{Kra05JPCM}.

\section{Discussion and conclusion}

In this paper, a mode-coupling theory for the slow dynamics of fluids
adsorbed in disordered porous solids made of spherical particles
frozen in random positions has been developed. Derived by properly
taking into account the peculiar structure of the correlations in
these systems and by including a contribution which had been forgotten
in a previous work \cite{Kra05PRL}, its equations are found to display
many appealing features. For instance, they show universality, in the
sense that they do not contain any explicit reference to the precise
nature of the random environment in which the fluid evolves. Also,
they compare favorably with previous mode-coupling equations derived
in other contexts, for the residual dynamics in the glassy phase of a
bulk fluid (Sec.~III) or for the equilibrium dynamics of a mean-field
spin-glass in a random magnetic field (Appendix
\ref{app.pspin}). Thus, from a formal point of view, the theory
appears rather satisfactory.

Nevertheless, a few difficulties remain.  First, there is the fact
that, in the limit of vanishing fluid-fluid interactions, the present
theory does not coincide with the MCT which can be derived by assuming
from the start that there are no such interactions. Second, there is
the divergence of the memory kernel for small wavevectors and the
resulting spurious long time anomalies
\cite{Got81PMB,Leu83PRAb,HofFreFraUNPUB}. We do not believe that these
issues are really harmful, even if their handling requires ad hoc
approximations, but they are worth stressing, since their solutions
would probably teach us something on the nature of the approximations
underlying the MCT scheme and on possible extensions of the
theory. For instance, it has been argued by Leutheusser that the
inclusion of vertex corrections within a kinetic theory approach would
solve the second problem, but no operational scheme was proposed for
this calculation \cite{Leu83PRAb}.

The numerical solution of the MCT equations for two simple
fluid-matrix models leads to a variety of transition scenarios, which
are either discontinuous for dilute matrices or continuous for dense
matrices. Depending on the model, in the intermediate region where the
nature of the transition changes, degenerate or genuine higher-order
singularities and glass-glass transition lines are found.

Another remarkable prediction of the theory is the possibility of a
reentry phenomenon for high matrix densities above the localization
threshold, which has been interpreted as the signature of a
decorrelation process induced by fluid-fluid collisions.  

Before going further, one should note that, strictly speaking, this
prediction of the theory cannot be correct in the case of hard core
fluid-matrix interactions. Indeed, as recently confirmed by extensive
computer simulations of the Lorentz gas \cite{HofFraFre06PRL}, the
localization transition is driven by the percolation transition of the
matrix void space, i.e., localization occurs because, above a certain
critical matrix density $\phi_m^*$, the void space only consists of
finite disconnected domains.  In such a scenario, it is obvious that,
whatever the fluid density, it is impossible to have an ergodic system
above $\phi_m^*$, since fluid-fluid interactions will never change the
geometry of the matrix.  This contradiction between the percolation
theory and the MCT clearly raises the issue of the relation between
the two approaches, for which we propose the following simple
argument. The MCT applied to the problem of the diffusion-localization
transition attempts to capture the onset of percolation in an indirect
way, by giving an account of the increasingly correlated nature of the
fluid-matrix collision events as the threshold is approached. In this
respect, one should note that none of the static structure functions
on which it is based does show a sensitivity to the phenomenon of
percolation.  Following Leutheusser, the theory works at the level of
a self-consistent treatment of ring collision processes
\cite{Leu83PRAb}. This turns out to be enough to predict a
diffusion-localization transition, but, clearly, the infinite
sequences of correlated collisions which would really reflect the
permanent trapping of the fluid particle in a finite domain above the
percolation threshold are missing. From this incomplete
characterization of the dynamical processes escorting the percolation
phenomenon, it results that the MCT diffusion-localization transition
is actually an ideal version of the true percolation transition, in
the usual sense that the MCT predicts ideal glass transitions, and
that the theory is not able to detect that, in an exact treatement,
the percolation threshold fixes an absolute limit to diffusive
behavior.  When fluid-fluid collisions come into play at finite fluid
densities, this leaves room for the prediction of a reentry phenomenon
in contradiction with percolation theory. We believe that it is for
the same reason that the MCT also misses the fact that, at any matrix
density, there is always a non-vanishing probability that particles
will be trapped in a finite domain disconnected from the rest of the
void space, so that the exploration of the available void space is
never completely ergodic \cite{KerMet83JPA}.

It is thus clear that the prediction of a reentrant behavior of the
ergodicity breaking transition line in the low fluid density domain
should not be taken too literally. In fact, a reasonable expectation
based on this finding is that, below the localization threshold, but
in the regime where transient trapping effects are important, there
might be an acceleration of the dynamics due to fluid-fluid
collisions.  Interestingly, such a behavior has already been observed
in a computer simulation study of a two dimensional lattice gas model
with fixed randomly placed scatterers \cite{Ole91JPA}. Indeed, it was
found that, starting from the zero fluid density limit, the diffusion
coefficient of a tagged particle first increases with the fluid
density. An interpretation in terms of a decorrelation process similar
to the one discussed in Sec. IV was then proposed and validated by
varying the parameters of the model.

As a possible source of the difficulties of the theory, one might
blame the fact that it works at the level of disorder averaged
quantities. Indeed, the procedure of averaging over disorder is
equivalent to an averaging over the volume of a macroscopic system, an
operation in which many microscopic details of the statics and
dynamics become blurred. This might confer a mean-field character to
the theory, where the contribution of the matrix would actually be
taken into account at the level of a diffuse effective localizing
potential, with a possible loss of important local constraints.
Unfortunately, this is a necessary step in order to develop a theory
which is comparable in complexity with the one for bulk systems, since
it allows one to consider the system as homogeneous.  Some progress
has recently been made on the application of the MCT to inhomogeneous
situations \cite{BirBouMyiRei06PRL}, so it should be possible to relax
the condition of homogeneity in order to study the dynamics of fluids
confined in pores of simple geometry which are often preferred in
simulation works, but there is no doubt that the more complex
wavevector dependence of the resulting theory will make it harder to
obtain numerical solutions of the equations.

Moreover, beyond this purely technical aspect, the present formulation
in terms of disorder averaged quantities has a practical interest as
well. Indeed, many real porous media are disordered and most
experimental techniques measure quantities which are averaged over the
volume of a macroscopic sample and thus equivalent to the disorder
averages considered by the MCT. So, the current theoretical setup
seems well suited for direct comparisons with experimental results.
For molecular dynamics simulations, however, since rather small
systems are usually considered, it might be necessary to explicitly
perform the disorder average over a representative sample of matrix
configurations before a comparison with the theory can be done.

Altogether, in spite of the above merely technical issues, we believe
that the present mode-coupling theory represents a valuable step
towards a better understanding of the slow dynamics of confined
glassforming liquids. Indeed, it remains rather simple and, since it
is a microscopic approach, it allows one to study in detail the effect
on the dynamics of changes in the different ingredients of a
fluid-matrix model (fluid-fluid and fluid-matrix interactions,
structure of the matrix). Thus, this provides us with a tool to
efficiently and thoroughly explore the phenomenology of dynamics in
confinement.  This is clearly illustrated by our findings for two very
simple systems with pure hard core interactions, which already display
new and nontrivial glass transition scenarios.  Then remains the
question of the validation of the theoretical predictions. Because the
model of the QA mixture is quite simple and the theory makes detailed
predictions, molecular dynamics studies should be able to give
clear-cut answers. The presently available results look rather
encouraging, but more simulation work is definitely needed.

\acknowledgments It is a pleasure to thank G. Tarjus, W. G{\"o}tze,
and W. Kob, for useful comments, and F. H{\"o}fling for a valuable
discussion and the communication of unpublished results.

\appendix
\section{Replica Ornstein-Zernike equations}
\label{app.oz}
For reference, we quote the replica OZ equations relating the various
pair correlation functions in QA binary mixtures
\cite{GivSte92JCP,LomGivSteWeiLev93PRE,GivSte94PA,RosTarSte94JCP}.
They read, in Fourier space,
\begin{subequations}
\begin{align}
\hat{h}^{mm}_q & = \hat{c}^{mm}_q + n_m \hat{c}^{mm}_q \hat{h}^{mm}_q,
\\
\hat{h}^{fm}_q & = \hat{c}^{fm}_q + n_m \hat{c}^{fm}_q \hat{h}^{mm}_q +
n_f \hat{c}^{c}_q \hat{h}^{fm}_q, \\
\hat{h}^{ff}_q & = \hat{c}^{ff}_q + n_m \hat{c}^{fm}_q \hat{h}^{fm}_q +
n_f \hat{c}^{ff}_q \hat{h}^{ff}_q - n_f \hat{c}^b_q \hat{h}^b_q, \\
\hat{h}^c_q & = \hat{c}^c_q + n_f \hat{c}^c_q \hat{h}^c_q,
\end{align}
\end{subequations}
with $\hat{c}^{ff}_q = \hat{c}^c_q + \hat{c}^b_q$ and $\hat{h}^{ff}_q
= \hat{h}^c_q + \hat{h}^b_q$. As usual, $h$ and $c$ denote total and
direct correlation functions, respectively. $\hat{f}$ denotes the
Fourier transform of $f$ and the superscripts have the same meaning as
for the structure factors (see Sec. II).

Using the relations
\begin{subequations}
\begin{align}
S^{mm}_q & = 1 + n_m \hat{h}^{mm}_q, \\
S^{fm}_q & = \sqrt{n_f n_m} \hat{h}^{fm}_q, \\
S^{c}_q & = 1 + n_f \hat{h}^{c}_q, \\
S^{b}_q & = n_f \hat{h}^{b}_q,
\end{align}
\end{subequations}
(remember that $S^{ff}_q=S^{c}_q + S^{b}_q$, hence $S^{ff}_q= 1 + n_f
\hat{h}^{ff}_q$) the OZ equations can be formally solved for the
structure factors, leading to
\begin{subequations}
\begin{align}
S^{mm}_q & = \frac{1}{1 - n_m \hat{c}^{mm}_q}, \\
S^{fm}_q & =  \frac{\sqrt{n_f n_m} \hat{c}^{fm}_q}{( 1 - n_m
\hat{c}^{mm}_q ) ( 1 - n_f \hat{c}^{c}_q )}, \\
S^{c}_q & = \frac{1}{1 - n_f \hat{c}^{c}_q}, \label{OZconn} \\
S^{b}_q & = n_f \left[\hat{c}^{b}_q + n_m \frac{(\hat{c}^{fm}_q)^2}{1-
n_m \hat{c}^{mm}_q}\right] \frac{1}{(1 - n_f \hat{c}^{c}_q)^2}.
\end{align}
\end{subequations}
\null

\section{Derivation of the mode-coupling memory kernel}
\label{app.derivation}
In this Appendix, the derivation of the mode-coupling part of the
memory kernel is outlined. This calculation is a direct extension of
the one for bulk systems which is described in its most minute details
in Ref.~\cite{LesHouches}.

The memory function in Eq.~\eqref{eqlangevin} is defined as
\begin{equation}\label{defkern}
\Omega_{q}^2 m_q(t)= \frac{\overline{\langle
R_\mathbf{q}(t)\,R_\mathbf{-q}\rangle}}{N_f m k_B T},
\end{equation}
where $R_\mathbf{q}(t) =\exp[i \mathcal{Q}_1 \mathcal{L} \mathcal{Q}_1
t] i \mathcal{Q}_1 \mathcal{L} g^f_\mathbf{q}$ is the projected random
force obtained from the longitudinal momentum density fluctuation
$g^f_\mathbf{q}(t)$ \cite{simpleliquids}. $\mathcal{L}$ is the
Liouville operator of the system and $\mathcal{Q}_1$ is the
complementary operator of the projector $\mathcal{P}_1$ which projects
any dynamical variable onto the subspace spanned by
$\delta\rho^f_\mathbf{q}$ and $g^f_\mathbf{q}$.

The calculation of the mode-coupling part of the kernel amounts to
replacing $R_\mathbf{q}$ in Eq.~\eqref{defkern} by its projection
$\mathcal{P}_2R_\mathbf{q}$ onto the subspace spanned by
$B_{\mathbf{q,k}}=\delta\rho^f_\mathbf{k} \delta\rho^f_\mathbf{q-k}$,
$C^{(1)}_{\mathbf{q,k}}=\delta\rho^f_\mathbf{k} \rho^m_\mathbf{q-k}$,
and $C^{(2)}_{\mathbf{q,k}}=\delta\rho^f_\mathbf{k} \langle
\rho^f_\mathbf{q-k} \rangle$, where we introduce the projection
operator $\mathcal{P}_2$ such that
\begin{widetext}
\begin{equation}
\mathcal{P}_2R_\mathbf{q}= {\sum_{\mathbf{k}}}' \overline{\langle
R_\mathbf{q} B_{\mathbf{-q,-k}}\rangle} G_{\mathbf{q},\mathbf{k}}
B_{\mathbf{q,k}} + \sum_{\mathbf{k}} \sum_{l,l'}\overline{\langle
R_\mathbf{q} C^{(l)}_{\mathbf{-q,-k}}\rangle}
H^{(ll')}_{\mathbf{q},\mathbf{k}} C^{(l')}_{\mathbf{q,k}}.
\end{equation}
\end{widetext}
In this expression, we have anticipated that, within the mode-coupling
approximation, $\mathcal{P}_2$ is diagonal in $\mathbf{k}$ and the
subspaces spanned by the $B$s and $C$s are orthogonal.  The prime in
the first sum indicates that, to avoid double-counting, the
wavevectors are assumed to be ordered somehow and the sum is
restricted to $\mathbf{k}<\mathbf{q-k}$. $G_{\mathbf{q},\mathbf{k}}$
and $H^{(ll')}_{\mathbf{q},\mathbf{k}}$ are normalization matrices
insuring that $\mathcal{P}_2B_\mathbf{q,k}=B_\mathbf{q,k}$ and
$\mathcal{P}_2C^{(l)}_{\mathbf{q,k}}=C^{(l)}_{\mathbf{q,k}}$.
 
In the course of this calculation, four-point density correlation
functions are generated. Thus, in order to eventually obtain closed
dynamical equations, a factorization approximation is needed to
express them as products of two-point density correlation
functions. We follow the usual mode-coupling prescription and find
that, within this approximation, the only non-vanishing four-point
functions are given by
\begin{subequations}
\begin{align}
\overline{\langle e^{i\mathcal{Q}_1\mathcal{L}\mathcal{Q}_1t}
B_{\mathbf{q,k}} B_{\mathbf{-q,\mathbf{-k}}}\rangle} & \simeq N_f^2
S^{c}_k S^{c}_{|\mathbf{q-k}|}\phi_{k}(t) \phi_{|\mathbf{q-k}|}(t),
\label{factbb} \\
\overline{\langle e^{i\mathcal{Q}_1\mathcal{L}\mathcal{Q}_1t}
C^{(1)}_{\mathbf{q,k}} C^{(1)}_{\mathbf{-q,\mathbf{-k}}}\rangle} &
\simeq N_f N_m S^{c}_k S^{mm}_{|\mathbf{q-k}|} \phi_{k}(t), \\
\overline{\langle e^{i\mathcal{Q}_1\mathcal{L}\mathcal{Q}_1t}
C^{(2)}_{\mathbf{q,k}} C^{(2)}_{\mathbf{-q,\mathbf{-k}}}\rangle} &
\simeq N_f^2 S^{c}_k S^{b}_{|\mathbf{q-k}|} \phi_{k}(t), \\
\overline{\langle e^{i\mathcal{Q}_1\mathcal{L}\mathcal{Q}_1t}
C^{(1)}_{\mathbf{q,k}} C^{(2)}_{\mathbf{-q,\mathbf{-k}}}\rangle} &
\simeq N_f \sqrt{N_f N_m} S^{c}_k S^{fm}_{|\mathbf{q-k}|} \phi_{k}(t).
\end{align}
\end{subequations}

Specializing to $t=0$ and computing $\mathcal{P}_2 B_\mathbf{q,k}$ and
$\mathcal{P}_2 C^{(l)}_{\mathbf{q,k}}$, it results that
\begin{subequations}
\begin{align}
G_{\mathbf{q},\mathbf{k}} = & \left[ N_f^2 S^{c}_k
S^{c}_{|\mathbf{q-k}|} \right]^{-1}, \\
H^{(11)}_{\mathbf{q},\mathbf{k}} = & \left[ N_f N_m S^{c}_k
\right]^{-1} \frac{S^{b}_{|\mathbf{q-k}|}}{S^{mm}_{|\mathbf{q-k}|}
S^{b}_{|\mathbf{q-k}|} - (S^{fm}_{|\mathbf{q-k}|})^2}, \\
H^{(22)}_{\mathbf{q},\mathbf{k}} = & \left[ N_f^2 S^{c}_k \right]^{-1}
\frac{S^{mm}_{|\mathbf{q-k}|}}{S^{mm}_{|\mathbf{q-k}|}
S^{b}_{|\mathbf{q-k}|} - (S^{fm}_{|\mathbf{q-k}|})^2}, \\
H^{(12)}_{\mathbf{q},\mathbf{k}} = & - \left[ N_f \sqrt{N_f N_m}
S^{c}_k \right]^{-1}
\frac{S^{fm}_{|\mathbf{q-k}|}}{S^{mm}_{|\mathbf{q-k}|}
S^{b}_{|\mathbf{q-k}|} - (S^{fm}_{|\mathbf{q-k}|})^2}.
\end{align}
\end{subequations}

It remains to express $\overline{\langle R_\mathbf{q}
B_{\mathbf{-q,-k}} \rangle} = \overline{\langle
i\mathcal{Q}_1\mathcal{L}g^f_\mathbf{q} B_{\mathbf{-q,-k}}\rangle}$
and $\overline{\langle R_\mathbf{q} C^{(l)}_{\mathbf{-q,-k}} \rangle}
= \overline{\langle i\mathcal{Q}_1\mathcal{L}g^f_\mathbf{q}
C^{(l)}_{\mathbf{-q,-k}}\rangle}$. First, using Yvon's theorem, one
finds
\begin{subequations}
\begin{align}
\overline{\langle i\mathcal{L}g^f_\mathbf{q}
B_{\mathbf{-q,-k}}\rangle} & =i q k_B T N_f \left[
\frac{\mathbf{q}\cdot \mathbf{k}}{q^2}
S^{c}_{|\mathbf{q-k}|}+\frac{\mathbf{q}\cdot(\mathbf{q-k})}{q^2}
S^{c}_k\right],\\
\overline{\langle i\mathcal{L}g^f_\mathbf{q}
C^{(1)}_{\mathbf{-q,-k}}\rangle} & =i q k_B T \sqrt{N_f N_m}
\frac{\mathbf{q}\cdot \mathbf{k}}{q^2} S^{fm}_{|\mathbf{q-k}|},\\
\overline{\langle i\mathcal{L}g^f_\mathbf{q}
C^{(2)}_{\mathbf{-q,-k}}\rangle} & = i q k_B T N_f
\frac{\mathbf{q}\cdot \mathbf{k}}{q^2} S^{b}_{|\mathbf{q-k}|}.
\end{align}
\end{subequations}
Then, with the help of the extension of the convolution approximation
\cite{JacFee62RMP} to QA systems,
\begin{subequations}
\begin{align}
\overline{\langle \delta\rho^f_{\mathbf{q}}
\delta\rho^{f}_{\mathbf{-k}} \delta\rho^{f}_{\mathbf{-q+k}} \rangle}&=
N_f S^{c}_q S^{c}_k S^{c}_{|\mathbf{q-k}|},\label{convconn}\\
\overline{\langle \delta\rho^f_{\mathbf{q}}
\delta\rho^{f}_{\mathbf{-k}} \rho^{m}_{\mathbf{-q+k}}\rangle}&=
\sqrt{N_f N_m} S^{c}_q S^{c}_k S^{fm}_{|\mathbf{q-k}|},\\
\overline{\langle \delta\rho^f_{\mathbf{q}}
\delta\rho^{f}_{\mathbf{-k}} \langle \rho^{f}_{\mathbf{-q+k}} \rangle
\rangle}&= N_f S^{c}_q S^{c}_k S^{b}_{|\mathbf{q-k}|},
\end{align}
\end{subequations}
it comes
\begin{subequations}
\begin{align}
\overline{\langle i\mathcal{P}_1\mathcal{L}g^f_\mathbf{q}
B_{\mathbf{-q,-k}}\rangle} & = i q k_B T N_f S^{c}_k
S^{c}_{|\mathbf{q-k}|}, \\
\overline{\langle i\mathcal{P}_1\mathcal{L}g^f_\mathbf{q}
C^{(1)}_{\mathbf{-q,-k}}\rangle} & = i q k_B T \sqrt{N_f N_m} S^{c}_k
S^{fm}_{|\mathbf{q-k}|}, \\
\overline{\langle i\mathcal{P}_1\mathcal{L}g^f_\mathbf{q}
C^{(2)}_{\mathbf{-q,-k}}\rangle} & = i q k_B T N_f S^{c}_k
S^{b}_{|\mathbf{q-k}|}.
\end{align}
\end{subequations}
The desired results are obtained by substracting the matching
equations.

We might now complete the explicit calculation of
$\mathcal{P}_2R_\mathbf{q}$. Indeed, if we define
\begin{align}
v_\mathbf{q,k} & = \overline{\langle R_\mathbf{q}
B_{\mathbf{-q,-k}}\rangle} G_{\mathbf{q},\mathbf{k}},\\
w^{(1)}_\mathbf{q,k} & = \overline{\langle R_\mathbf{q}
C^{(1)}_{\mathbf{-q,-k}}\rangle} H^{(11)}_{\mathbf{q},\mathbf{k}} +
\overline{\langle R_\mathbf{q} C^{(2)}_{\mathbf{-q,-k}}\rangle}
H^{(21)}_{\mathbf{q},\mathbf{k}},\\
w^{(2)}_\mathbf{q,k} & = \overline{\langle R_\mathbf{q}
C^{(1)}_{\mathbf{-q,-k}}\rangle} H^{(12)}_{\mathbf{q},\mathbf{k}} +
\overline{\langle R_\mathbf{q} C^{(2)}_{\mathbf{-q,-k}}\rangle}
H^{(22)}_{\mathbf{q},\mathbf{k}},
\end{align}
such that
\begin{equation}
\mathcal{P}_2R_\mathbf{q}= {\sum_{\mathbf{k}}}' v_\mathbf{q,k}
B_{\mathbf{q,k}} + \sum_{\mathbf{k}} w^{(1)}_\mathbf{q,k}
C^{(1)}_{\mathbf{q,k}} + w^{(2)}_\mathbf{q,k} C^{(2)}_{\mathbf{q,k}},
\end{equation}
the above results immediately lead to 
\begin{align}
v_\mathbf{q,k} & = \frac{i q k_B T}{N_f} \left[ \frac{\mathbf{q}\cdot
\mathbf{k}}{q^2} \frac{1}{S^{c}_k} +
\frac{\mathbf{q}\cdot(\mathbf{q-k})}{q^2}
\frac{1}{S^{c}_{|\mathbf{q-k}|}} - 1 \right],\\
w^{(1)}_\mathbf{q,k} & = 0, \\
w^{(2)}_\mathbf{q,k} & = \frac{i q k_B T}{N_f} \left[
\frac{\mathbf{q}\cdot \mathbf{k}}{q^2} \frac{1}{S^{c}_k} - 1 \right].
\end{align}
Note the vanishing of the fluid-matrix contribution.

Eventually, injecting the resulting expression of
$\mathcal{P}_2R_\mathbf{q}$ into
\begin{equation}
\Omega_{q}^2 m^{(\text{MC})}_q(t) = \frac{\overline{\langle
\mathcal{P}_2R_\mathbf{q}(t)\,\mathcal{P}_2R_\mathbf{-q}\rangle}}{N_f
m k_B T},
\end{equation}
one obtains Eqs.~\eqref{kernel} and \eqref{vertices} after a few
elementary steps. 

Exactly the same calculation can be done with subsets of the above
three quadratic variables. One then easily demonstrates that, if only
$\delta\rho^f_\mathbf{k} \delta\rho^f_\mathbf{q-k}$ and
$\delta\rho^f_\mathbf{k} \langle \rho^f_\mathbf{q-k} \rangle$ are
considered, the same equations are obtained, while, if one works with
$\delta\rho^f_\mathbf{k} \delta\rho^f_\mathbf{q-k}$ and
$\delta\rho^f_\mathbf{k} \rho^m_\mathbf{q-k}$, the equations of
Ref.~\cite{Kra05PRL} are reproduced, which reduce to those of
Refs.~\cite{GotLeuYip81PRA,Leu83PRA,Sza04EL} in the zero fluid density
limit.

\section{Mean-field spin-glass in a random magnetic field}
\label{app.pspin}
A fruitful source of new theoretical developments on the physics of
glassy systems has been the finding that the MCT provides an exact
description of the equilibrium dynamics of a certain class of
mean-field spin-glass models with multispin interactions
\cite{KirThi87PRL,CriHorSom93ZPB,BouCugKurMez96PA}.  In this Appendix,
based on Ref.~\cite{CiuCri00EL} (the interested reader is referred to
this paper for details), we show that the equations describing the
dynamics of such a mean-field spin-glass model in a random magnetic
field reproduce the structure of those of the MCT for a fluid in a
random environment. Thus, the correspondence between the two
approaches still holds in the presence of an external source of
disorder.

We consider the fully connected mean-field spherical spin-glass model
with three spin interactions in a random magnetic field, whose
Hamiltonian is
\begin{equation}
H_J[\mathbf{s}] = - \sum_{1\le i<j<k\le N}J^{(3)}_{ijk} s_i s_j s_k -
  \sum_{1\le i\le N} J^{(1)}_{i} s_i,
\end{equation}
where the spins $s_i$ are $N$ real variables subject to the constraint
$\sum_{i=1}^N s_i^2=N$, and the random coupling constants
$J^{(3)}_{ijk}$ and fields $J^{(1)}_i$ are uncorrelated zero mean
Gaussian variables with variances $3 J_3^2/N^2$ and $J_1^2/2$,
respectively. In zero field ($J_1=0$), this model is known to generate
the quadratic memory kernel introduced in Refs.~\cite{BenGotSjo84JPC}
and \cite{Leu84PRA} as a one wavevector approximation to the bulk MCT
functional. Thus, it is especially suitable for our purpose, since it
displays nonlinearities of the same degree as in the MCT.

In the presence of a random magnetic field, just like QA mixtures are
characterized by static frozen density fluctuations, the present
spin-glass model is characterized by static frozen local
magnetizations $\langle s_i \rangle$ which vanish when the average
over disorder (both on couplings and fields) is
performed. Accordingly, we might define $\delta s_i(t) = s_i(t) -
\langle s_i \rangle$ and introduce three correlation functions, which
are the analogues of $S^{b}_q$, $S^{c}_q$, and $\phi_q(t)$. They are,
respectively, the two overlap functions
\begin{equation}
q_b = \frac1N \sum_{i=1}^N \overline{\langle s_i \rangle^2 }
\end{equation}
and 
\begin{equation}
q_c = \frac1N \sum_{i=1}^N \overline{\langle (\delta s_i)^2 \rangle },
\end{equation}
which obey $q_b+q_c=1$, and the normalized connected spin correlation
function,
\begin{equation}
\phi(t) = \frac{1}{N q_c} \sum_{i=1}^N \overline{\langle \delta s_i(t)
\delta s_i(0) \rangle}.
\end{equation}

The exact solution of the Langevin dynamics of the model using
standard methods \cite{CiuCri00EL} provides us with an equation for
the time evolution of $\phi(t)$ which reads, for a high enough
temperature $T$,
\begin{subequations}\label{pspin1}
\begin{equation}
\dot{\phi}(t) + \frac{1}{q_c} \phi(t) + \frac{1}{q_c} \int_0^t d\tau
m(t-\tau) \dot{\phi}(\tau)=0,
\end{equation}
with
\begin{equation}
m(t) = \frac12 q_c \left[\frac{3 J_3^2}{T^2}\right] q_c^2 \phi(t)^2 +
  q_c \left[\frac{3 J_3^2}{T^2}\right] q_c q_b \phi(t).
\end{equation}
\end{subequations}
A direct comparison of these equations with those of Sec.~II
immediately shows that both sets of equations exactly have the same
formal structure. In particular, beyond the simple fact that the
memory functions are similar polynomials of the dynamical correlation
functions, the same parametrization in terms of connected and
disconnected static correlations is found.

We conclude with two short remarks. First, as for QA mixtures, we note
that the external random field does not enter explicitly in the above
equations (no $J_1$ is present). Because the model is exactly soluble,
it is easy to understand how this happens. In fact, the details of the
random field are only needed for the computation of the static
correlations, through the equality
\begin{equation}
\frac{J_1^2}{2T^2} + \frac{3 J_3^2}{2T^2} q_b^2 =
\frac{q_b}{(1-q_b)^2}.
\end{equation}
Once this calculation is done, they can be forgotten. The same is
probably true for fluids in random environments and this supports the
assumption that the MCT equations for QA systems actually have a wider
domain of applicability, provided they are expressed in terms of fluid
quantities only.

Second, it can be shown that Eq.~\eqref{pspin1} also describes the
residual relaxation of the present spin-glass model in zero field,
when the system is equilibrated in its low temperature phase and the
total spin correlation function has a nonvanishing infinite time limit
$q$ \cite{BarBurMez96JPA}. Then, $q_b$ and $q_c$ simply have to be
replaced by $q$ and $1-q$. Thus, in the present model, the difference
discussed in Sec.~III between self-induced glassiness and the effect
of a quenched random field does not seem to exist. This could have
been anticipated, since by construction, fully connected models are
unable to capture phenomena which only show up in the wavevector
dependence of the coupling constants in fluid systems.


\begin{thebibliography}{99}
\bibitem{QuantumNote} For instance, the theory of
  Ref.~\cite{ThaNei96PRB} amounts to setting $\hat{c}^{c}_q\equiv0$ in
  Eq.~\eqref{vone}, but not in Eq.~\eqref{vtwo}.
\bibitem{IdealGas} For an ideal gas in any type of statistically
homogeneous random environment, one finds that
$c^{c}(r)=h^{c}(r)\equiv 0$ and that $c^{b}(r)$ and $h^{b}(r)$ are
independent of $n_f$.
\bibitem{notePY} The possible values of $\Delta$ are limited by the
method of solution of the OZ/PY equations on a real space grid of
equally spaced points with a step size commensurate to the hard core
diameter $d$. For our calculation, the present choice of $\Delta$
offers the closest match with the value $\Delta=0.4/d$ used in
Ref.~\cite{FraFucGotMaySin97PRE}, where analytic expressions for the
structure functions were readily available.
\end{thebibliography}
\end{document}